\documentclass[useAMS,usenatbib]{mn2e}

\usepackage{times}  % to get nice fonts for PDF
\usepackage{listings}
\usepackage{graphics}
\usepackage{subfigure}
\usepackage{graphicx}
\usepackage{amssymb}
\usepackage{hyperref}
\usepackage{aas_macros}

\def\simprop{ \lower .75ex \hbox{$\sim$} \llap{\raise .27ex \hbox{$\propto$}} }

\title[Which galaxies dominate the neutral gas content?]{
Which galaxies dominate the neutral gas content of the Universe?}
\author[Claudia del P. Lagos et al.]{
\parbox[t]{\textwidth}{
\vspace{-1.0cm}
C. D. P. Lagos$^{1}$,
C. M. Baugh$^{2}$,
M. A. Zwaan$^{1}$,
C. G. Lacey$^{2}$,
V. Gonzalez-Perez$^{3}$,
C. Power$^{4}$,
A. M. Swinbank$^{2}$,
E. van Kampen$^{1}$
}
\vspace*{6pt} \\
$^{1}$European Southern Observatory, Karl-Schwarzschild-Strasse 2, 85748, Garching, Germany.\\
$^{2}$Institute for Computational Cosmology, Department of Physics,
University of Durham, South Road, Durham, DH1 3LE, UK.\\
$^{3}$Centre de Physique des Particules de Marseille, Aix-Marseille Universit\'e, CNRS/IN2P3, Marseille, France.\\
$^{4}$International Centre for Radio Astronomy Research, University of Western Australia, 35 Stirling Highway, Crawley, WA 6009, Australia 
\vspace*{-0.5cm}}

\begin{document}

%\date{Accepted ???. Received ???; in original form ???}

\pagerange{\pageref{firstpage}--\pageref{lastpage}} \pubyear{2011}

\maketitle

\label{firstpage}

\begin{abstract}
We study the contribution of galaxies with different properties to the
global densities of star formation rate (SFR), atomic (HI) and molecular hydrogen (H$_2$) as a function of redshift.
We use the
{\texttt{GALFORM}} model of galaxy formation, which is set
in the $\Lambda$CDM framework. This model includes a self-consistent
calculation of the SFR, which depends on the H$_2$
 content of galaxies.
 The predicted SFR density and how much of this is contributed by
galaxies with different stellar masses and infrared luminosities are in
agreement with observations. The model predicts a modest evolution of the
HI density at $z<3$, which is also in agreement with the observations. The HI density 
is predicted to be always dominated by galaxies with SFR$<1 M_{\odot}\,\rm yr^{-1}$.
This contrasts with the H$_2$ density, which is predicted to be dominated by 
 galaxies with SFR$>10M_{\odot}\,\rm yr^{-1}$ at $z>1$.
Current high-redshift galaxy surveys
are limited to detect carbon monoxide in galaxies with ${\rm SFR}\gtrsim 30 M_{\odot}\,\rm yr^{-1}$, which
in our model make up, at most, $20$\% of the H$_2$ in the universe.
In terms of stellar mass,
the predicted H$_2$ density is dominated by massive galaxies, $M_{\rm stellar}>10^{10}\,M_{\odot}$,
 while the HI density is dominated by low mass galaxies, $M_{\rm stellar}<10^{9}\,M_{\odot}$.
 In the context of upcoming neutral gas surveys, we suggest that the faint nature of the galaxies dominating the HI content of the Universe
will hamper the identification of optical counterparts, while for
H$_2$, we expect follow up observations of molecular emission lines
of already existing galaxy catalogues to be able to uncover the H$_2$ density of the Universe.
\end{abstract}


\section{Introduction}

Observations of local and high-redshift galaxies indicate that stars form from 
 molecular gas, {albeit with a low efficiency} (e.g. \citealt{Bigiel08}; 
\citealt{Genzel10}). On scales of kilo-parsecs, this fact is observed as a  
close to linear correlation between the star formation rate (SFR) surface density 
and the surface density of molecular hydrogen (H$_2$) (\citealt{Bigiel08}; \citealt{Bigiel10}; \citealt{Leroy08}; \citealt{Schruba11}).
The observational and theoretical 
evidence gathered point to a scenario in which gas cools down and becomes molecular before collapsing and fragmenting 
 to form stars, {with the small efficiency explained as being due to self-regulation as the results of star formation 
at the level of giant molecular clouds}. Therefore, the presence of neutral gas by itself does not guarantee 
star formation (see \citealt{Kennicutt12} for a recent review). This emphasises the importance 
of H$_2$ as the {main tracer of the dense regions where stars form}. 
Regarding the atomic hydrogen (HI), studies of local Universe galaxies show a 
strong correlation between the HI mass and the stellar mass, indicating that {the evolution of the two quantities is related} 
  (e.g. \citealt{Catinella10}; \citealt{Cortese11}; \citealt{Huang12}).
In addition, the morphology of the 
atomic hydrogen in galaxies is closely 
related to baryonic processes such as gas accretion and outflows in galaxies 
(\citealt{Fraternali02}; \citealt{Oosterloo07}; \citealt{Boomsma08}).
Observations have shown that the presence of stellar driven outflows depends on 
the star formation rate density (e.g. 
\citealt{Chen10}; \citealt{Newman12}). \citet{Lagos13} and \citet{Creasey12}, using hydrodynamical modelling of the growth of supernovae 
driven bubbles in the ISM of galaxies, show that the fundamental property setting the outflow rate 
is the gas surface density, as it affects both the SFR, which sets the energy input from supernovae, and the time bubbles take 
to escape the galaxy disk.

All of the evidence above points to the fact that a key step in understanding galaxy formation is 
 the observation of multiple gas phases in the ISM and their relation to the presence of SF.
In particular, observations of atomic and molecular hydrogen, which make up most of the  
mass in the interstellar medium (ISM) of galaxies, are key to understanding how SF proceeds 
 and how gas abundances can be linked to accretion and outflow of gas. 
Only through tracing both molecular and atomic gas components at the same time 
 as the SFR and stellar mass in galaxies will we be able to develop a better understanding of 
the processes of SF and feedback and to put strong constraints on galaxy formation simulations. 

Currently, observational constraints on the 
HI and H$_2$ contents of galaxies are available 
for large local samples, and for increasingly large samples at high-redshift.
In the case of HI, accurate measurements of the $21$~cm emission in large 
surveys of local galaxies have been presented by \citet{Zwaan05} using the HI Parkes 
All-Sky Survey (HIPASS; \citealt{Meyer04}) and by \citet{Martin10} 
using the Arecibo Legacy Fast ALFA Survey (ALFALFA; \citealt{Giovanelli05}). 
The global HI mass density at $z=0$ has been estimated from these surveys to be in the range 
$\Omega_{\rm HI}=3.6-4.2 \times 10^{-4}$, with little evolution up 
to $z\approx 5$ (e.g. \citealt{Peroux03}; \citealt{Noterdaeme09}).
This lack of evolution is fundamentally different 
from the evolution inferred in the SFR density, which shows a strong increase from $z\approx 0$ to $z\approx 2-3$, 
followed by a slow decline to higher redshifts (e.g. \citealt{Madau96}; \citealt{Lilly96}; \citealt{Hopkins06}). 

A plausible explanation for the different evolution seen in the SFR and HI densities  
is that HI is not the direct fuel of SF
but instead this gas still needs to cool down further to form stars. This indicates that 
 in order to better understand SF and galaxy evolution, good observations of the 
 denser gas in the ISM, i.e. H$_2$, are required. 
Molecular hydrogen is commonly traced by the $^{12}\rm CO$ (hereafter `CO') molecule, 
which is the second most abundant molecule after H$_2$ and is easily excited. 
However, direct CO detections 
are very scarce and complete samples are limited to the local Universe.  
\citet{Keres03} reported the first attempt to derive the 
local $\rm CO(1-0)$ luminosity function  
from which they inferred the H$_2$ mass function  and the local 
$\Omega_{\rm H_2}=1.1\pm 0.4\times 10^{-4} \, h^{-1}$, adopting a 
Milky-Way type $\rm CO(1-0)$-H$_2$ conversion factor. 
It has not yet been possible 
to estimate the cosmic H$_2$ abundance at high redshift. 
Observational samples that 
detect CO at high-redshifts are limited to moderately and highly star-forming galaxies with 
SFR$\gtrsim 20\, M_{\odot}\, \rm yr^{-1}$ (\citealt{Tacconi13}; \citealt{Carilli13}). 
To estimate a density of H$_2$ from these galaxies is difficult given the selection effects and volume corrections. 
Recently, \citet{Berta13} inferred  
 the molecular gas mass function at high-redshift by using the ratio between the UV to mid-IR emission and its empirical relation 
 to the molecular gas mass \citep{Nordon13}. However, these estimates are subject to the uncertain extrapolation 
of empirical relations to galaxies and redshifts where they have not been measured. 
 For these reasons the construction of 
representative samples of H$_2$ in galaxies is still an unfinished task that is 
crucial to our understanding of galaxy formation.

From the theoretical point of view, more sophisticated treatments of SF and its relation to 
the abundance of HI and H$_2$ have only recently started to be explored in ab-initio 
cosmological galaxy formation simulations (e.g. \citealt{Fu10}; \citealt{Cook10};  
\citealt{Lagos10}; \citealt{Kuhlen12}; \citealt{Duffy12b}; \citealt{Popping13}), as well as in 
 simulations of individual galaxies (e.g. \citealt{Gnedin09}; \citealt{Pelupessy09}; \citealt{Kim211}; \citealt{MacLow12}). 
The improved modelling of the ISM and SF in galaxies that has been applied 
in cosmological galaxy formation simulations has allowed a better understanding of the 
neutral content of galaxies; for instance, of the effect of H$_2$ self-shielding on 
 the column density distribution of HI (\citealt{Altay10}; \citealt{Duffy12b}), the 
lack of evolution of the global HI density (\citealt{Lagos11}; \citealt{Dave13}), and 
the increasing molecular gas-fractions of galaxies with redshift (\citealt{Obreschkow09c}; \citealt{Lagos12}; \citealt{Fu12}). 

The issue motivating this paper is the contribution of star-forming galaxies of different properties to the 
atomic and molecular gas content of the Universe. {Note that this question is similar to asking 
about the steepness of the faint-end of the gas and stellar mass functions; i.e. the steeper the mass function, the 
larger the contribution from low-mass galaxies to the total density of mass in the Universe.}
 This exercise is important for two reasons. 
First, blind CO surveys are very challenging and expensive. For instance, the Atamaca Large Millimeter Array can 
 easily detect galaxies at high-redshift (e.g. \citealt{Vieira13}; \citealt{Hodge13}), 
but has the downside of having a very small field of view ($<1$~arcmin), which is not ideal for 
large surveys. A possible solution is to 
 follow-up samples of galaxies selected
by SFR or stellar mass in molecular emission, which can give insight into 
the relation between SF and the different gas phases in the ISM.
Second, because a key science driver of all large HI programs is  
to understand how the HI content relates to multi-wavelength galaxy properties (see for instance the accepted 
proposals of the ASKAP HI All-Sky Survey \footnote{{\tt http://www.atnf.csiro.au/research/WALLABY/proposal.html}}, WALLABY, and 
the Deep Investigation of Neutral Gas Origins survey\footnote{\tt http://askap.org/dingo}, DINGO, \citealt{Johnston08}). DINGO 
will investigate this directly as it {overlaps} with the Galaxy And Mass Assembly multi-wavelength survey \citep{Driver09}. 

In this paper we bring theoretical models a step closer to the observations 
and examine how star-forming galaxies {of different properties} contribute to the densities of SFR, atomic and molecular hydrogen, 
to inform future neutral gas galaxy surveys about 
the expected contribution of star-forming galaxies to the HI and H$_2$ contents of the Universe.
We also seek to understand how far we currently are 
from uncovering most of the gas 
in galaxies in the Universe.
For this study we use three flavours of the semi-analytical model {\texttt{GALFORM}} 
in a $\Lambda$CDM
cosmology (\citealt{Cole00}),
namely those of \citet{Lagos12}, \cite{Gonzalez-Perez13}, and Lacey et al. (2014, in prep.). 
The three models include the improved treatment of SF implemented by \citet{Lagos10}. This extension 
 explicitly splits the hydrogen content of the ISM 
of galaxies into HI and H$_2$. The advantage of using three different flavours of  {\texttt{GALFORM}}
 is the ability to characterise the robustness of the trends found. The outputs of the three models 
shown in this paper will be publicly available from the Millennium 
database\footnote{\tt http://gavo.mpa-garching.mpg.de/Millennium} (detailed in Appendix~\ref{Resolution}).

This paper is organised as follows. In $\S 2$ we present the galaxy formation model 
and describe the realistic SF model used. We also describe the main differences between the three 
models that use this SF law and the dark matter simulations used.
 In $\S 3$ we compare the model predictions with observations of the evolution of the 
SFR, HI and H$_2$ densities, and the SF activity in normal and highly star-forming galaxies. None of the 
observations in $\S 3$ were used to constrain model parameters, and {therefore provide independent verifications of the predictions 
of the models}. 
 Given the success of our model predictions, we investigate the contribution from star-forming galaxies 
 to the densities of HI and H$_2$ in $\S 4$ and discuss 
the physical drivers of the predictions.
In $\S 5$ we discuss the implications of our predictions for the next generation of HI and H$_2$ surveys and present 
the main conclusions in $\S 6$. 

\section{Modelling the two-phase cold gas in galaxies}\label{modelssec}

In this section, 
 we briefly describe the main aspects of the {\texttt{GALFORM}} 
semi-analytical model of galaxy formation and evolution \citep{Cole00}, 
focusing on the key features of the three flavours adopted in this study, 
which are described in 
Lagos et al. (2012; hereafter `Lagos12'), Gonzalez-Perez et al. (2013; hereafter `Gonzalez-Perez14') and
Lacey et al. (2014; hereafter `Lacey14').

The three {\tt GALFORM} models above  
take into account the main physical processes
that shape the formation and evolution of galaxies. These are: (i) the collapse
and
merging of dark matter (DM) halos, (ii) the shock-heating and radiative cooling
of gas inside
DM halos, leading to the formation of galactic disks, (iii) quiescent star
formation in galaxy disks, (iv) feedback
from supernovae (SNe), from active galactic nucleus (AGN) 
heating and from photo-ionization of the
inter-galactic medium (IGM), (v) chemical 
enrichment of stars and gas, and (vi) galaxy mergers driven by
dynamical friction within common DM halos which can trigger bursts of SF,
and lead to the formation of spheroids (for a review of these
ingredients see \citealt{Baugh06}; \citealt{Benson10b}). 
Galaxy luminosities are computed from the predicted star formation and
chemical enrichment histories using a stellar population synthesis model (see \citealt{Gonzalez-Perez13}). 

The three flavours of {\tt GALFORM} used in this study adopt the same SF law, which is a key process affecting the 
evolution of the gas content of galaxies. In $\S$~\ref{SFlaw} we describe this choice of SF law and how this connects 
to the two-phase ISM, in $\S$~\ref{Models} we describe the differences
between the three flavours of {\tt GALFORM} and in $\S$\ref{Cosmos} we briefly 
describe the dark matter simulations used for this study and the cosmological parameters adopted.

\subsection{Interstellar medium gas phases and the star formation law}\label{SFlaw}

The three flavours of {\tt GALFORM} used in this study 
 adopt the SF law developed in Lagos et al. (2011a, hereafter `L11'), 
in which the atomic and molecular 
phases of the neutral hydrogen in the ISM are explicitly distinguished.
L11 found that the SF law that gives the best agreement with 
the observations without the need for fine tunning is the 
empirical SF law of \citet{Blitz06}. 
Given that the SF law has been well constrained in spiral and dwarf galaxies in the 
local Universe, L11 decided to implement this molecular-based SF law only in the quiescent SF mode (SF due to gas 
accretion onto the disk), keeping the original prescription of \citet{Cole00} 
for starbursts (driven by galaxy mergers and 
global disk instabilities). We provide more details below.

{\it Quiescent Star Formation.} The empirical SF law of Blitz \& Rosolowsky has the form,
\begin{equation}
\Sigma_{\rm SFR} = \nu_{\rm SF} \,\rm f_{\rm mol} \, \Sigma_{\rm gas},
\label{Eq.SFR}
\end{equation}
\noindent where $\Sigma_{\rm SFR}$ and $\Sigma_{\rm gas}$ are the surface
densities of SFR and total cold gas mass (molecular and atomic), respectively, 
$\nu_{\rm SF}$ is the inverse of the SF 
timescale for the molecular gas and $\rm f_{\rm mol}=\Sigma_{\rm mol}/\Sigma_{\rm gas}$ is the
molecular to total gas mass surface density ratio. The molecular and total gas 
contents include the contribution from helium, while HI and H$_2$ only include  
hydrogen (which in total corresponds 
to a fraction $X_{\rm H}=0.74$ of the overall cold gas mass). 
The ratio $\rm f_{\rm mol}$ depends on
the internal hydrostatic pressure as $\Sigma_{\rm H_2}/\Sigma_{\rm HI}=\rm f_{\rm mol}/(f_{\rm mol}-1)=
(P_{\rm
ext}/P_{0})^{\alpha}$.
{To calculate $\rm P_{\rm
ext}$, we use the approximation from \citet{Elmegreen89}, in which the pressure 
depends on the surface density of gas and stars. We give the values of the parameters involved in this SF law
in $\S$~\ref{Models}.

{\it Starbusts.} In starbursts the SF timescale is proportional to the bulge dynamical
timescale above a minimum floor value 
and involves the whole cold gas content of the galaxy, $\rm SFR={\it M}_{\rm cold}/\tau_{\rm SF}$ 
(see \citealt{Granato00} and \citealt{Lacey08} for details). The SF timescale is defined as

\begin{equation}
\tau_{\rm SF}=\rm max(\tau_{\rm min},f_{\rm dyn}\tau_{\rm dyn}), 
\label{SFlawSB}
\end{equation}

\noindent where $\tau_{\rm dyn}$ is the bulge dynamical timescale,  $\tau_{\rm min}$ is a minimum duration 
of starbursts and $f_{\rm dyn}$ is a parameter. We give the values of the parameters involved in this SF law 
in $\S$~\ref{Models}.

Throughout this work 
we assume that in starbursts, the cold gas content is fully molecular, $\rm f_{\rm mol}=1$. 
Note that this is similar to assuming that the relation between the ratio 
$\Sigma_{\rm H_2}/\Sigma_{\rm HI}$ and $P_{\rm ext}$ 
holds in starbursts 
given that large gas and stellar densities lead to 
 $\rm f_{\rm mol}\approx1$. Throughout the paper we refer to galaxies going through a starburst, which are 
 driven by galaxy mergers or disk instabilities, as `starbursts'.

Another key component of the ISM in galaxies is dust. To compute the extinction of starlight 
we adopt the method described 
 in \citet{Lacey11} and \citet{Gonzalez-Perez12}, which uses the results of 
the physical radiative transfer model of \citet{Ferrara99} to calculate the dust attenuation at different wavelength.
The dust model assumes
a two-phase interstellar medium, with star-forming clouds embedded in 
a diffuse medium. The total mass of dust is predicted by
{\texttt{GALFORM}} self-consistently from the cold gas mass and
metallicity, assuming a dust-to-gas ratio which is proportional to the
gas metallicity. The radius of the diffuse dust component is
assumed to be equal to 
the half-mass radius of the disk, in the case of quiescent SF, or the bulge, in the case of starbursts.
 With this model, we can estimate the attenuation in the wavelength range from the far-ultraviolet (FUV) 
to the near-infrared (IR), including emission lines. 
In {\tt GALFORM} the extinction for emission lines is calculated 
as due to the diffuse dust component only, so it is the same as the extinction of older stars 
(i.e. outside their birth clouds) at the same wavelength. 
We define the total IR luminosity
 as the total luminosity emitted by interstellar
dust, free from contamination by starlight, which approximates to  
the integral over the rest-frame wavelength range 8$-$1000~$\mu$m. 

\subsection{Differences between the Lagos12, Gonzalez-Perez14 and Lacey14 models}\label{Models} 

The Lagos12 model is a development of the model originally described in \citet{Bower06}, which was the 
first variant of {\tt GALFORM} to include AGN feedback as the mechanism suppressing gas cooling 
in massive halos. The Lagos12 model assumes a universal initial mass function (IMF), 
the \citet{Kennicutt83} IMF\footnote{The distribution of the masses of stars 
formed follows ${\rm d}N(m)/{\rm d\, ln}\,m \propto m^{-x}$, where $N$ is the number of stars of mass $m$ formed, 
 and $x$ is the IMF slope. For a \citet{Kennicutt83} IMF, $x=1.5$ for masses in the range $1\,M_{\odot}\le m\le 100\,M_{\odot}$ and 
$x=0.4$ for masses $m< 1\,M_{\odot}$.}.
Lagos12 extend the model of Bower et al. by including the self-consistent SF law 
described in $\S$~\ref{SFlaw}, and adopting $\nu_{\rm SF}=0.5\,\rm Gyr^{-1}$, 
$\rm log(P_{0}/k_{\rm B} [\rm cm^{-3} K])=4.23$, where 
$\rm k_{\rm B}$ is is Boltzmann's constant, and $\alpha=0.8$, 
which correspond to the volume of the parameters reported by \citet{Leroy08} for local spiral 
and dwarf galaxies. 
This choice of SF law greatly reduces the parameter space of the model and also 
extends its predictive power by directly modelling the atomic and molecular hydrogen content of galaxies. 
All of the subsequent models that make use of the same SF law have also the ability to predict
the HI and H$_2$ gas contents of galaxies. 
Lagos12 adopt longer duration starbursts (i.e. larger $f_{\rm dyn}$) 
compared to Bower et al. to improve the agreement with the observed luminosity 
function in the rest-frame ultraviolet (UV) at high redshifts. Lagos12 adopts $\tau_{\rm min}=100\, \rm Myr$ and 
$f_{\rm dyn}=50$ in Eq.~\ref{SFlawSB}. The Lagos12 model was developed in the Millennium simulation, which 
assumed a WMAP1 cosmology \citep{Spergel03}. 

The Gonzalez-Perez14 model updated the Lagos12 model to the WMAP7 cosmology \citep{Komatsu11} and vary few model 
parameters to 
help recover the agreement between the model predictions and the observed evolution 
of both the UV and $K$-band luminosity functions. These changes include 
a slightly shorter starburst duration, i.e.  $\tau_{\rm min}=50\, \rm Myr$ and $f_{\rm dyn}=10$, and 
 weaker supernovae feedback. 

The Lacey14 models is also developed on the WMAP7 cosmology, as the Gonzalez-Perez14 model, but it differs 
from both the Lagos12 and Gonzalez-Perez14 models in that it adopts a bimodal IMF. The IMF describing 
SF in disks (i.e. the quiescent mode) is the same as the universal IMF in the other two models, but 
a top-heavy IMF is adopted for starbursts (i.e. with an IMF slope $x=1$). This is inspired by \citet{Baugh05} who used a bimodal 
IMF to recover the agreement between the models predictions and observations of 
the number counts and redshift distribution of submillimeter galaxies.
 We notice, however, that Baugh et al. adopted a more top-heavy IMF for starbursts with 
 $x=0$. 
The stellar population synthesis model for Lacey14 is also different. 
While both Lagos12 and Gonzalez-Perez14 use \citet{Bruzual03}, the Lacey14 model makes use of 
\citet{Maraston05}.
Another key difference between the Lacey14 model and the other two {\tt GALFORM} flavours considered here, is that 
 Lacey14 adopt a slightly larger value of the SF timescale, $\nu_{\rm SF}=1.2\,\rm Gyr^{-1}$, still within 
the range allowed {by the most recent observation compilation of \citet{Bigiel11}}, making SF more efficient. 

\subsection{Halo merger trees and cosmological parameters}\label{Cosmos}

{\texttt{GALFORM}} uses
the formation histories of DM halos as a starting point
to model galaxy formation (see \citealt{Cole00}).
In this paper we use Monte-Carlo generated merger histories from \citet{Parkinson08}.
We adopt a minimum halo mass of 
$M_{\rm halo,min}=5 \times 10^{8} h^{-1} M_{\odot}$ and {model a representative sample of dark matter halos 
such that the predictions presented for the mass functions and integrated 
distribution functions are robust.}
The minimum halo mass chosen enables us to predict cold gas mass structures down to the current observational 
limits (i.e. $M_{\rm HI}\approx 10^{6} M_{\odot}$, \citealt{Martin10}). At
higher redshifts, this minimum halo mass is scaled with redshift as $(1+z)^{-3}$ 
to roughly track the evolution of the break in the halo mass function, so that
we simulate objects with a comparable range of space densities at each redshift. 
This allows us to follow a representative sample of dark matter halos. 
We aim to study in detail the HI and H$_2$ contents of galaxies in the universe. 
%{The volume chosen is the result of a compromise between the minimum halo mass and the minimization of cosmic variance. 
%This volume is $3$ times higher than the Millennium-II simulation, which \citet{Guo11} show to have very little 
%cosmic variance at $M^{*}$.}
To achieve this,
the use of Monte-Carlo trees is compulsory as the range of halo masses needed to do such a study is much larger 
than the range in currently available $N$-body simulations, such as the Millennium \citep{Springel05} and 
the Millennium-II\footnote{Data from the Millennium/Millennium-II simulation is available on a relational database 
accessible from {\tt http://galaxy-catalogue.dur.ac.uk:8080/Millennium}.}
 \citep{Boylan-Kolchin09}. In Appendix~\ref{Resolution}, we discuss in detail the effect of resolution on the neutral 
gas content of galaxies and show that the chosen minimum halo mass for this study is sufficient to achieve our goal. 

Lagos12 adopt the
cosmological parameters of the Millennium N-body
simulation \citep{Springel05}: $\Omega_{\rm m}=\Omega_{\rm DM}+\Omega_{\rm baryons}=0.25$ (with a
baryon fraction of $0.18$), $\Omega_{\Lambda}=0.75$, $\sigma_{8}=0.9$
and $h=0.73$, while Gonzalez-Perez14 and Lacey14 adopt WMAP7 parameters: 
$\Omega_{\rm m}=\Omega_{\rm DM}+\Omega_{\rm baryons}=0.272$ (with a
baryon fraction of $0.167$), $\Omega_{\Lambda}=0.728$, $\sigma_{8}=0.81$
and $h=0.704$. 

\section{The evolution of SFR, atomic and molecular gas densities}\label{cosmicevo}

In this section we explore 
the predictions for the evolution of the H$\alpha$ and UV
luminosity functions, and the global densities 
of SFR, HI and H$_2$ ($\S$~\ref{Sec:Models}). The aim of this is to establish how well the models 
describe the star-forming galaxy population at different redshifts and 
 if the overall predicted gas content of the universe is in agreement with observational estimates. 
 We then focus in one model, the one that best reproduces the observations, to 
 study the contribution to the SFR density from galaxies of different stellar masses and 
IR luminosities in $\S$~\ref{Sec:Local} and $\S$~\ref{Sec:SBs}.

\subsection{Global densities of SFR and neutral gas: models vs. observations}\label{Sec:Models} 

\subsubsection{The H$\alpha$ and UV luminosity functions}
\begin{figure}
\begin{center}
\includegraphics[trim = 0.7mm 2mm 1mm 1mm,clip,width=0.47\textwidth]{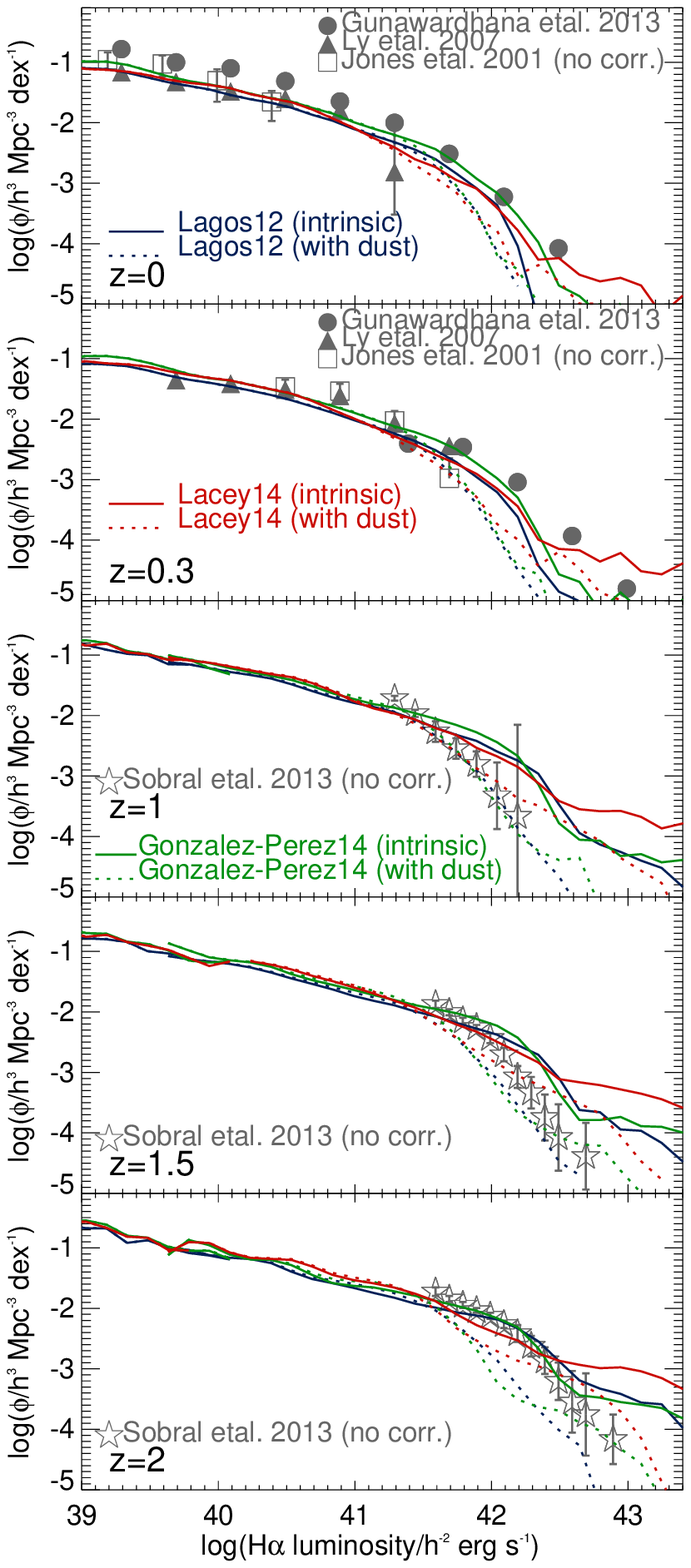}
\caption{The H$\alpha$ luminosity function, intrinsic and attenuated by dust, 
at different redshifts (indicated at the lower-left corner of each panel), 
for the Lagos12, Lacey14 and 
Gonzalez-Perez14 models, as labelled. 
 Observational 
estimates of the H$\alpha$ luminosity function from \citet{Jones01}, \citet{Ly07} and \citet{Gunawardhana13} are shown 
in the top two panels, and from \citet{Sobral13} in the bottom three panels using symbols, as labelled.
Filled symbols represent observational data that has been 
corrected for dust extinction (i.e. comparable to the model intrinsic luminosity function), and open symbols 
represent the observed luminosity function (i.e. comparable to the model extinct luminosity function). {Note that we express 
densities and luminosities in comoving units for both the observations and the models. To transform these units to physical units 
the reader needs to evaluate $h$ with the value given in $\S$~\ref{Cosmos}.}}
\label{HalphaLF}
\end{center}
\end{figure}

\begin{figure}
\begin{center}
\includegraphics[trim = 0.7mm 1mm 1mm 1mm,clip,width=0.47\textwidth]{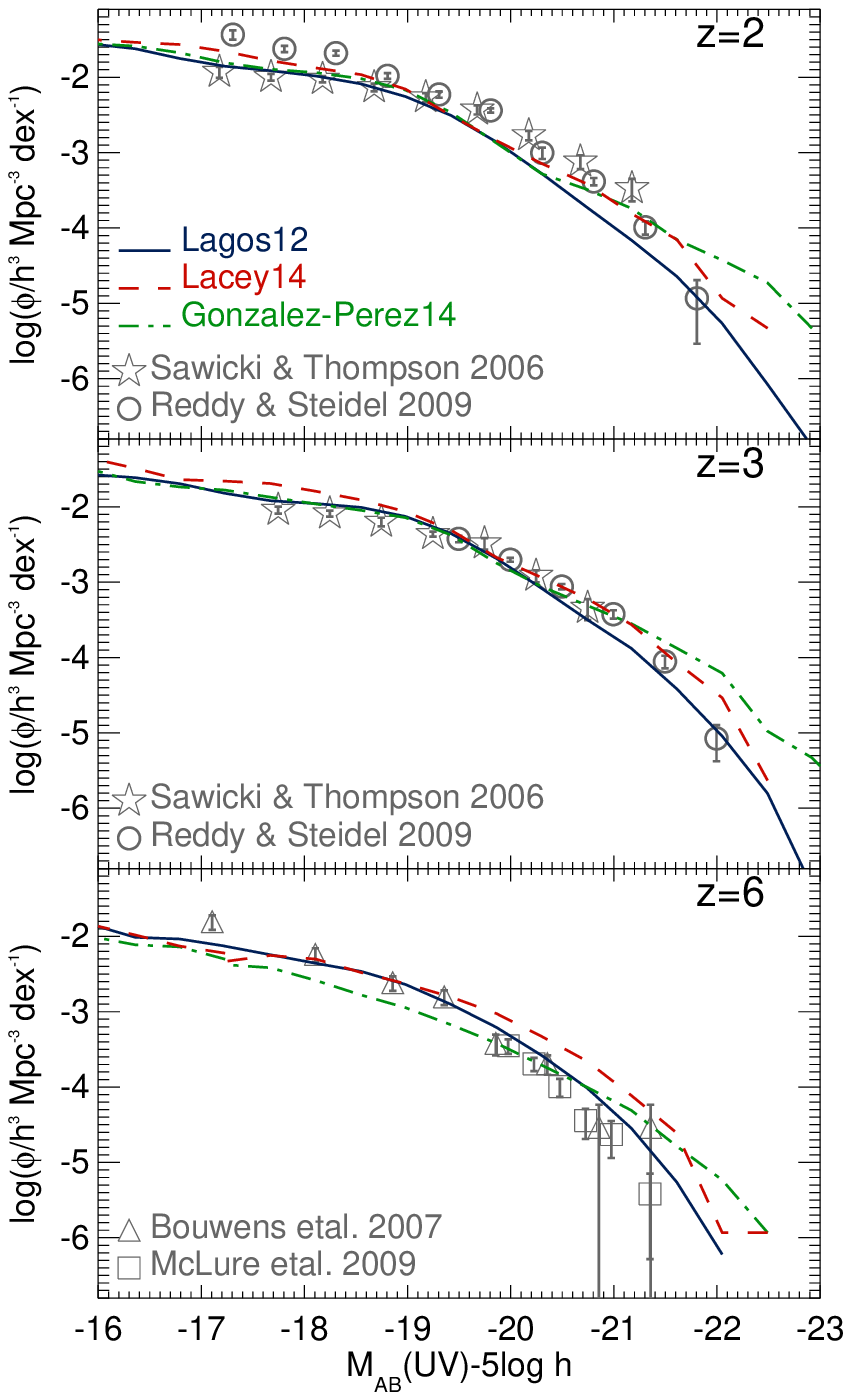}
\caption{The rest-frame UV(1500\AA) luminosity function at $z=2$ (top panel), $z=3$ (middle panel) 
and $z=6$ (bottom panel) for the Lagos12, Lacey14 and 
Gonzalez-Perez14 models, as labelled. Symbols represent the 
observational estimates from \citet{Sawicki06},  \citet{Bouwens07}, \citet{Reddy09} and 
\citet{McLure09}, as labelled. The predictions shown for the models already include the effects of dust extinction 
and should be compared to the observed luminosity functions without any correction for extinction.{The units of density here 
are as in Fig.~\ref{HalphaLF}. However, to convert the magnitudes from the adopted units to $M_{\rm AB}(\rm UV)$ the reader 
needs to add $5\rm log\,h$ to the quantity in the $x$-axis, where $h$ has the value showed in $\S$~\ref{Cosmos}}.}
\label{UVLF}
\end{center}
\end{figure}

A fair comparison between the observations and the model predictions for the SFR can be made by directly contrasting the 
observed H$\alpha$ and UV luminosity functions with the predicted ones, 
which correspond to two of the most popular SFR tracers. Fig.~\ref{HalphaLF} shows the 
intrinsic H$\alpha$ luminosity function from $z\approx 0$ to $z\approx 1.5$  for the three models and 
observations. The observations have been separated into those which present the $H\alpha$ luminosity function including 
a correction for dust attenuation (filled symbols), and those which do not (open symbols). 
For the three models we show both the intrinsic and dust extincted $H\alpha$ luminosity functions.
The intrinsic luminosity functions predicted by the 
models should be compared to the filled symbols. 

The three models have predictions in reasonable agreement with the observations, within 
the errorbars. At $z=0$ and $z=0.3$, the models predict $H\alpha$ luminosity functions in broad agreement with the 
observations. 
At $z=1$, $z=1.5$ and $z=2$ we only show direct observations of the H$\alpha$ luminosity function from \citet{Sobral13}, who 
present the largest galaxy samples to date, greatly exceeding the number of galaxies and observed sky area from previous works
 (i.e. $14,011$ emitters in $2.2\,\rm deg^2$).
 These observations are directly comparable to the model predictions that include dust.
The three models are consistent with the $z\approx 1$ observations, but are 
below the observations at $z\approx 1.5$ and $z\approx 2$ at the faint-end by a factor of $\approx 2-3$. 
This may be due either to a low number of highly star-forming galaxies in the models or 
 to the dust extinction modelling. 
The former would change the intrinsic H$\alpha$ luminosity function, while the latter would affect only the extincted 
H$\alpha$ emission. We expect both to contribute to some extent.
Differences between the three models in Fig.~\ref{HalphaLF} are mainly seen at the bright-end where 
the errorbars on the observations are the largest.
In general, our dust model 
predicts that dust attenuation is luminosity dependent, which is a natural consequence 
of the different gas metallicities in our model galaxies. The faint-end is expected to be only slightly 
affected by dust, but bright $H\alpha$ galaxies can be highly attenuated by dust. However, the predicted extinctions 
for $z\gtrsim 1$ are typically smaller than that assumed in observations ($A_{\rm H\alpha}=1$~mag). 
The Lacey14 model predicts higher number densities of luminous H$\alpha$ galaxies, with intrinsic 
$L_{\rm H\alpha}>5\times 10^{42}\,h^{-2}\,\rm erg\,s^{-1}$, than the other two models. This is due the predominance 
of starbursts at these high luminosities and the top-heavy IMF adopted by Lacey14 in bursts. 
Note that 
the three models predict that the normalisation and break of the H$\alpha$ luminosity function increase 
with increasing redshift between $z=0$ and $z=1$, while at $z>1$, the evolution is mainly in the number density of 
bright galaxies. This overall evolution agrees well with that observed. 

At high redshifts, a popular way of studying SF in galaxies is to use the rest-frame FUV emission as a tracer. 
We compare the predicted rest-frame FUV luminosity function in the three flavours of {\tt GALFORM} to observations from $z=2$ to 
$z=6$ in Fig.~\ref{UVLF}. The observed luminosity function corresponds to the rest-frame $1500$~\AA~luminosity, without any attempt 
to correct for dust extinction. This is directly comparable to the model predictions for the rest-frame $1500$~\AA~luminosity, after dust 
has been included in the model (lines in Fig.~\ref{UVLF}). The predictions from the three flavours of {\tt GALFORM} 
agree well with the observations, even up to $z=6$. This is partially due to the assumed duration of starbursts. \citet{Lacey11} 
show that very short starbursts, such as the ones assumed in \citet{Bower06}, 
result in the bright-end of the UV luminosity function being largely over predicted. 
Longer starbursts allow the model to have more dusty starbursts and reproduce the break in the UV LF. 
The duration of the bursts is of about few $100$~Myr, and therefore in better agreement with the 
latest estimates of the duration from observational data \citep{Swinbank13}. 

\subsubsection{The cosmic density of SFR}

\begin{figure}
\begin{center}
\includegraphics[trim = 0.9mm 2.8mm 1mm 3.45mm,clip,width=0.48\textwidth]{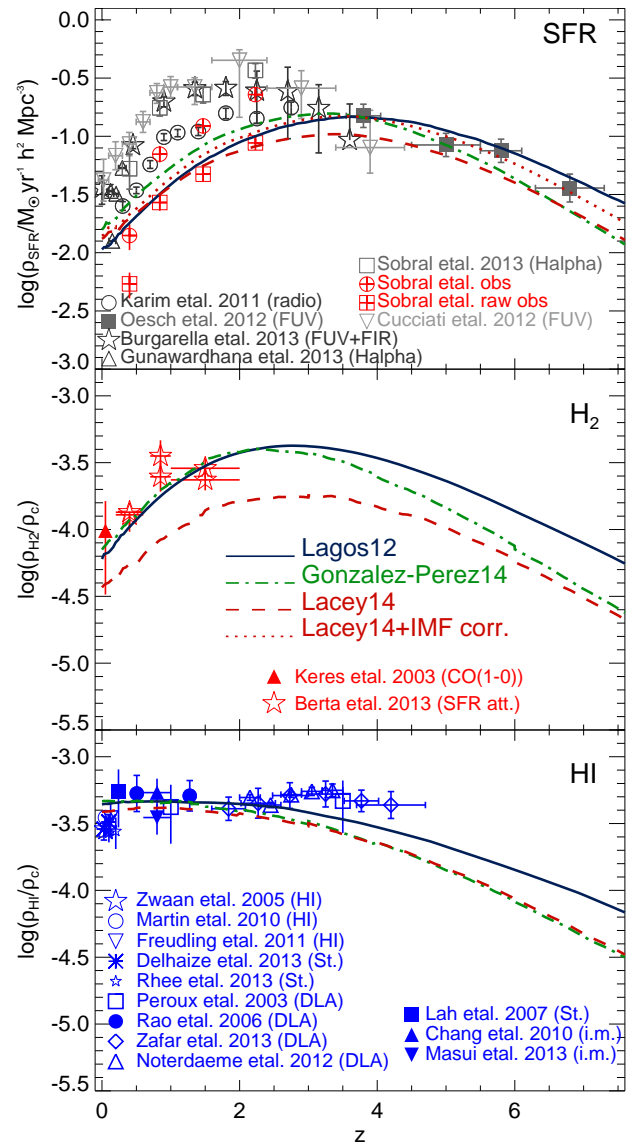}
\caption{{\it Top panel:} Evolution of the global density of the SFR in units of $M_{\odot}\, {\rm yr}^{-1}\, h^2\,{\rm Mpc}^{-3}$
 for the Lagos12, Lacey14 and Gonzalez-Perez14 models, as labelled.
For the Lacey14 model we also show the scaled $\rho_{\rm SFR}$ that would be inferred if a 
Kennicutt IMF was adopted as universal (dotted line; see text).
The grey squares and triangles  
correspond to the observational estimates of $\rho_{\rm SFR}$ from \citet{Sobral13} and \citet{Gunawardhana13}, respectively, using 
 H$\alpha$,  
grey circles to the estimates of \citet{Karim11}, using radio stacking, 
 inverted triangles and filled squares to estimates using the 1500\AA~emission corrected for dust extinction 
from \citet{Cucciati12} and \citet{Oesch12}, respectively, 
and stars to the estimates of \citet{Burgarella13} using FUV and FIR. {For Sobral et al. we also show their 
reported $\rho_{\rm SFR}$ integrated only over the range where they have measurements (without extrapolation; circles with inner crosses), 
and the latter without the dust correction (squares with inner crosses).} 
{\it Middle panel:} Evolution of the global density of H$_2$, in units of 
the critical density, for the same models.
Symbols show the $z\approx 0$ estimate of $\rho_{\rm H_2}$ from \citet{Keres03}
using the $\rm CO(1-0)$ luminosity function (filled triangle) and inferences at higher redshifts from 
\citet{Berta13} using the molecular masses inferred from the ratio between the UV to the IR emission (open stars).
{\it Bottom panel:} Evolution of the global density of HI, in units of  
the critical density, for the three models of the top panel.
Observations of $\rho_{\rm HI}$ from \citet{Zwaan05}, \citet{Martin10} and \citet{Freudling11} 
using 21~cm emission, \citet{Lah07}, \citet{Delhaize13} and \citet{Rhee13} from spectral stacking, \citet{Chang10} and 
\citet{Masui13} from intensity mapping, and \citet{Peroux03},
 \citet{Rao06}, \citet{Noterdaeme12} and \citet{Zafar13} from DLAs, are also shown using symbols, as labelled.}
\label{denstots}
\end{center}
\end{figure}

We are interested in connecting the evolution of the gas content of galaxies with that of the SFR. 
With this in mind, we show in Fig.~\ref{denstots} the evolution of the global  
 density of SFR,
HI and H$_2$ for the Lagos12, Lacey14 and Gonzalez-Perez14 models.
In the case of the SFR, the observational estimates have been corrected to our 
choice of IMF \citep{Kennicutt83} (see Appendix~\ref{IMFs} for the conversions).
{Note that we express the SFR density in comoving units as in Fig.~\ref{HalphaLF}.}
{The predictions of the three models 
are in reasonable agreement with the observational estimates at $z<0.3$ and at $z>2$, within the uncertainties, 
but significantly lower than the observational estimates in the redshift range $0.3<z<2$.}
The comparison between the model predictions and the observed density of SFR needs to be undertaken carefully 
as there are many systematic uncertainties which are not taken into account in the errorbars of the 
observations. One example is the
dust absorption correction applied when ultraviolet bands or emission lines are used to trace SFR (e.g. \citealt{Hopkins06}).
Typically, empirical dust corrections calibrated in the local Universe are extrapolated to high-redshifts (e.g.
\citealt{Calzetti07}). 
Another source of uncertainty is related to the extrapolation needed to correct the observed SFR density
to account for faint galaxies that are not detected. These two factors drive
most of the dispersion observed in Fig.~\ref{denstots} between the different estimates. 

{To help illustrate the impact of the extrapolation to fainter luminosities and the dust correction, we show in 
the top panel of Fig.~\ref{denstots} the SFR density from \citet{Sobral13} which includes both the dust correction and 
the extrapolation to faint galaxies (squares), the SFR density with the dust correction but integrated only over the range covered 
by the measurements (circles with inner crosses) and the latter but without the dust correction (squares with inner 
crosses). The overall effect of the dust correction is at most of $\approx 0.4$~dex and of the extrapolation can be an 
increase of up to $\approx 0.6$~dex. 
If the true slope of the H$\alpha$ LF is shallower and/or the dust correction is less dramatic than assumed in the observations 
the inferred $\rho_{\rm SFR}$ would easily move down closer to our predictions. Recently, \citet{Utomo14} show 
evidence that the dust corrections applied in most of the observations shown in the top panel of Fig.~\ref{denstots} are 
overestimated, leading to an overestimation of the SFRs.}
Other systematics are hidden in the adopted IMF. Although we scale the inferred observations to our adopted IMF,
this conversion is valid only in the case of a simple SF history, for instance a constant SFR, and a constant metallicity, 
at a certain stellar population age. This is
not the case for our simulated galaxies that undergo starbursts and gas accretion which can change the available gas
 content at any time \citep{Mitchell13}. 
 \citet{Oti-Floranes10} argue that the systematic uncertainty due to the IMF choice can be
up to a factor of $\approx 4$.
\citet{Wilkins12} show that the ratio between the SFR and the FUV luminosity 
is correlated with SFR, and that systematic errors relating to adopting a universal ratio can be a factor of $\approx 1.5$ or more.
A good example of the uncertainties in the observationally inferred $\rho_{\rm SFR}$ is the fact that 
 the three models predict a H$\alpha$ luminosity function  at $z\le 1$ in broad agreement with 
the observations (see Fig.~\ref{HalphaLF}), but a $\rho_{\rm SFR}$ that is a factor of $\approx 2-4$ lower 
than the H$\alpha$ inferred $\rho_{\rm SFR}$. 
 Thus, we consider the SFR density inferred from observations 
as a rough indicator of the SF activity as opposed to a stringent constrain to our model predictions. 
The fact that the three models have $\rho_{\rm SFR}$ at $z\approx 1$ below  
 the observations is not currently of great concern for two reasons: (i) given the difficulty of quantifying the effect of 
systematic biases on the observationally inferred $\rho_{\rm SFR}$, and (ii) 
because the predicted H$\alpha$ and UV luminosity functions 
are in reasonable agreement with observations. 

Overall, the three models predict $\rho_{\rm SFR}$ of a similar shape and normalisation. However, in the detail,  
we can distinguish two differences that are worth exploring: (i) the Lacey14 model 
predicts a lower peak of $\rho_{\rm SFR}$ compared to the other two models, and (ii) at $z>6$, there is an offset 
of a factor of $2-3$ between $\rho_{\rm SFR}$ predicted by the Lagos12 model and the other two models. 
Feature (i) is due to the adopted top-heavy IMF in starbursts in the Lacey14 model, which results in lower SFRs 
to drive the same level of FUV, IR and emission line 
fluxes compared to the choice of a Kennicutt IMF. In fact, we can calculate 
 the difference between the two different IMFs in the flux predicted for the different 
SFR tracers (see Appendix~\ref{IMFs}).
We can scale the contribution from starbursts up by 
$1.8$ to account for the different IMF; this SFR density would roughly correspond to the inferred one if a 
Kennicutt IMF was adopted as 
universal and UV emission was used as a tracer of SF.
This is shown by the dotted line in the top panel of Fig.~\ref{denstots}. This scaling accounts for most of
the difference between the Lacey14 and the other two models that adopt a universal Kennicutt IMF.
Feature (ii) is due to the different cosmologies adopted in the Lagos12 (WMAP1) and the other two models (WMAP7).
  The WMAP7 cosmology has a lower $\sigma_{8}$, which delays the collapse of the first structures,  
giving rise to the first galaxies at lower redshifts  
compared to the WMAP1 universe {(see \citealt{Gonzalez-Perez13} and \citealt{Guo13} for an analysis of the 
effect of these cosmologies on galaxy properties)}. 
Note that the difference in $\Omega_{\rm m}$ reported by 
WMAP1 and WMAP7 is irrelevant at these high redshifts as $\Omega_{\rm m}\approx 1$. 

\subsubsection{The cosmic densities of HI and H$_2$}

The middle and bottom panels of Fig.~\ref{denstots} 
shows that the atomic and molecular gas densities are predicted to have very different evolution in the 
three {\tt GALFORM} models.
{Note that we have converted all of the observations to our adopted cosmology ($\S$~\ref{Cosmos}).}
The predicted atomic gas density displays very little evolution at $z\lesssim 2.5$. At higher redshifts, 
 $\rho_{\rm HI}$ slowly decreases with increasing redshift. This behaviour is common in the three models.
Quantitatively, $\rho_{\rm HI}$ decreases by only $\approx 11$\% between $z=0$ and $z=2.5$ in the three models, 
while between $z=2.5$ and $z=6$ it 
decreases by $68$\% in the Lagos12 model, $82$\% in the Gonzalez-Perez14 model and 
$80$\% in the Lacey14 model. At $z\lesssim 3$, the predicted evolution of the 
three models agrees well with observations of $\rho_{\rm HI}$ that come either from direct HI detections at $z\approx 0-0.2$ or 
inferences from spectral stacking, intensity mapping and damped-Ly$\alpha$ systems (DLAs) at $z\gtrsim 0.2$. 
At $z>3$, the predictions for $\rho_{\rm HI}$ from the three models 
systematically deviate from the observed $\rho_{\rm HI}$. One reason for this is that 
we only model neutral gas inside galaxies, i.e. in the ISM, and we do not attempt to estimate the neutral fraction of the 
gas outside galaxies (i.e. in the IGM). This argument is supported by the results of hydrodynamical simulations which explicitly follow the 
neutral fraction in different environments. For instance, 
\citet{VanDeVoort12} and \citet{Dave13} show that the neutral content of the universe in their simulations 
is dominated by gas outside galaxies at $z\gtrsim 3$. 

The weak evolution of $\rho_{\rm HI}$ shown in Fig.~\ref{denstots} contrasts with the 
evolution of the molecular hydrogen global density, $\rho_{\rm H_2}$, which shows a steep increase between 
$z=0$ and $z=3$ followed by a slow decrease as redshift increases in the three models. Quantitatively, 
$\rho_{\rm H_2}$ increases by a factor $\approx 7$ between $z=0$ and $z=3$ in three models, followed by a decline of a factor of 
$\approx 3$ for the Lagos12 and Lacey14 models, and 
of $\approx 5$ for the Gonzalez-Perez14 model 
between $z=3$ and $z=6$. At $z=0$, the Lagos12 and the Gonzalez-Perez14 models 
predict a density of H$_2$ that is in good agreement with the inference of \citet{Keres03}, while the Lacey14 model 
predicts a H$_2$ density that is only marginally consistent with the observations. In Appendix~\ref{MFs} we show the 
predicted H$_2$ mass function in the three models compared to observations of Keres et al. and show that this tension 
between the Lacey14 model and the observationally inferred $\rho_{\rm H_2}$ rises from the lower number density of 
galaxies at the knee of the mass function. We warn the reader, however, that the sample of Keres et al. is not a blind CO survey 
and also makes use of a constant conversion factor between CO and H$_2$. Therefore it is unclear how much of this tension 
is real. (see \citealt{Lagos12} for an illustration of how the CO-H$_2$ ratio can vary.)

We have also included in the middle panel of 
Fig.~\ref{denstots} recent inferences of $\rho_{\rm H_2}$ from \citet{Berta13}. Berta et al. 
use the SFR attenuation (the ratio between the emission from the UV and from the mid-IR) to convert to molecular 
gas content using the empirical relation of \citet{Nordon13}.
 The two sets of symbols correspond to integrating the inferred H$_2$ mass function 
in the range derived by the observed SFRs and integrating down to H$_2$ masses of $10^7\,M_{\odot}$, which requires 
 extrapolation of the inferred 
H$_2$ mass function. These two sets of data serve as an indication 
of the uncertainties in $\rho_{\rm H_2}$, although no systematic 
effects are included in the errorbars. The Lagos12 and Gonzalez-Perez14 models are broadly consistent with the Berta et al. 
inference, while 
the Lacey14 model lies below.  
%This is not surprising as the observational data points were inferred from SFRs and are therefore subject to the 
%uncertainties described above for $\rho_{\rm SFR}$.
% This points to the need to use independent measurements of 
%$\rho_{\rm H_2}$ based on direct detection of molecular emission lines, or alternatively, 
%for accurate measurements of the dust content and gas metallicity. This would help to 
%constrain $\rho_{\rm H_2}$ independently of 
%$\rho_{\rm SFR}$. 
In $\S$~\ref{Sec:PhysDrivers}, we describe how the physical mechanisms included in the models interplay 
to drive the evolution of the 
three quantities shown in Fig.~\ref{denstots}.

The Lagos12, Lacey14 and Gonzalez-Perez14 models predict interesting differences, that can be tested in the future. 
However, with current observations of galaxies, we cannot distinguish between the models. 
Given that the Lagos12 model has been extensively explored in terms of 
 the gas abundance of galaxies (see for instance \citealt{Geach11} for the evolution of the molecular gas fractions,
 \citealt{Lagos12} for the evolution of the CO-FIR luminosity relation and luminosity functions,  
and \citealt{Kim13b} for an analysis of the HI mass function and clustering), we use this model in the next sections to 
explore the contribution from star-forming galaxies to the densities of SFR, HI and H$_2$. 
In $\S$~\ref{OtherMods}, we 
explore how robust these predictions are by comparing with the Lacey14 and Gonzalez-Perez14 models. 
However, we remark that 
the other two models give similar results to the Lagos12 model in the comparisons presented in $\S$~\ref{Sec:Local} and \ref{Sec:SBs}.  
 
\subsection{The evolution of normal star-forming galaxies}\label{Sec:Local}

\begin{figure}
\begin{center}
\includegraphics[width=0.49\textwidth]{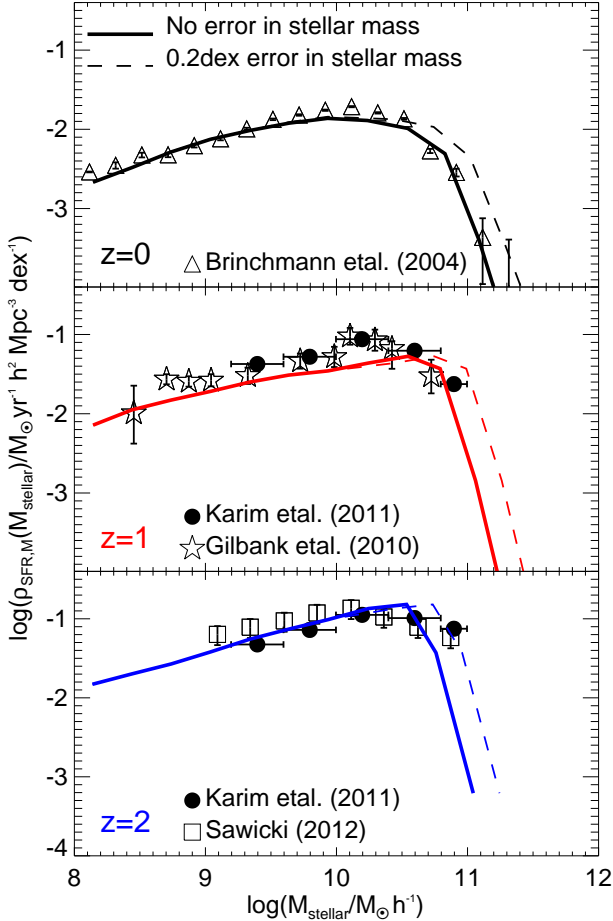}
\caption{Star formation rate density as a function of stellar mass for 
three different redshifts, as labelled, for the Lagos12 model. Solid lines 
show the model predictions and dashed lines show, for reference, the effect of including an error of $0.2$~dex in the stellar mass 
estimate (see \citealt{Mitchell13} for a complete discussion on the uncertainties in stellar mass estimates).
We show the observational results from \citet{Brinchmann04} at $z\approx 0.2$, 
\citet{Gilbank10} at $z\approx 1$, \citet{Karim11} at $z\approx 1$ and $z\approx 2$, and \citet{Sawicki12} at $z\approx 2$.
{Note that we express
densities and masses in comoving units for both the observations and the models. To transform these units to physical units
the reader needs to divide the mass by the value of $h$ given in $\S$~\ref{Cosmos}, and multiply the density by the $h$ value 
squared.}}
\label{denstots2}
\end{center}
\end{figure}

An important 
test of the model is the observed relation between the SFR density at fixed stellar mass, $\rho_{\rm SFR,M}$, 
and stellar mass at a given redshift. Several works have 
analysed this relation and concluded that it is very difficult for current semi-analytic models and simulations of galaxy formation 
to predict the right trends. The argument is that this relation is very sensitive to feedback mechanisms and how 
inflow and outflow compensate to each other (e.g. \citealt{Dutton09}; \citealt{Fontanot09}; 
\citealt{Lagos10}; \citealt{Dave11}; \citealt{Weinmann12}). 

Observationally, $\rho_{\rm SFR,M}$ is constructed using galaxies 
with detected SFRs (i.e. above a sensitivity limit in SFR). These galaxies are refer to as `star-forming', and lie on a tight 
relationship between the stellar mass and the SFR (\citealt{Schiminovich07}; \citealt{Noeske07}; \citealt{Elbaz07}), with 
 outliers typically being starbursts, which have disturbed dynamics (e.g. \citealt{Wuyts11}). 
However, in the model we include all the galaxies in a given stellar mass bin, even if they have low SFRs, 
  to calculate $\rho_{\rm SFR,M}$ given that passive galaxies, which 
lie below the sequence of star-forming galaxies, contribute very little to the global SFR (\citealt{Lagos10}). We show 
the $\rho_{\rm SFR,M}$-stellar mass distribution function at three different redshifts in Fig.~\ref{denstots2}. Observational data 
have been scaled to our choice of IMF using the scalings described in Appendix~\ref{IMFs}.

There are three interesting aspects of Fig.~\ref{denstots2}. First, the Lagos12 model predicts that the highest contribution 
to $\rho_{\rm SFR}$ comes from relatively massive galaxies and that this peak mass, 
$M_{\rm s}$, increases slightly with increasing redshift, 
from $M_{\rm s}\approx 2\times 10^{10}\, M_{\odot}\,h^{-1}$ at $z=0$ 
to $M_{\rm s}\approx 4\times 10^{10}\, M_{\odot}\,h^{-1}$ at $z=2$. The stellar mass corresponding to the peak of 
$\rho_{\rm SFR,M}$ is also consistent with the recent observational 
inferences of the $\rho_{\rm SFR,M}$ distribution function by \citet{Sobral13}.
In the model, 
this increase of a factor of $2$ is driven by the higher frequency of bright starbursts at high redshifts, which are also 
more frequent in massive galaxies, $M_{\rm stellar}>10^{10}\, M_{\odot}\,h^{-1}$ (see \citealt{Lagos10}).
Secondly, the 
drop in the contribution to $\rho_{\rm SFR}$ from galaxies with stellar masses above $M_{\rm s}$ becomes 
steeper with increasing redshift, which is related to the low number density of galaxies more massive than $M_{\rm s}$ 
at high-redshift; i.e. the break in the stellar mass function becomes smaller than $M_{\rm s}$ at $z\gtrsim 2$ \citep{Behroozi13}. 
Finally, the slope of the relation between $\rho_{\rm SFR,M}$ and $M_{\rm stellar}$ below $M_{\rm s}$ is very weakly dependent 
on redshift, but the normalisation increases with increasing redshift. Below $M_{\rm s}$, most of the 
galaxies lie on the SFR-stellar mass sequence of star-forming galaxies. Thus the slope of the $\rho_{\rm SFR,M}$-$M_{\rm stellar}$ 
is set by the slope of the SFR-stellar mass relation. \citet{Lagos10} show that this slope is sensitive to how quickly the gas 
that is expelled from the galaxy by stellar feedback reincorporates into the halo and becomes available for further cooling. 
Lagos et al. show that the model predicts a slope close to the observed one 
if this reincorporation timescale is short (i.e. of the order 
of a few halo dynamical timescales). The observations do not have yet the volumes and sensitivity limits needed to 
probe these three features, but with the current observations and the comparison with the model predictions 
we can conclusively say that the normalisation of the 
$\rho_{\rm SFR,M}$-$M_{\rm stellar}$ relation increases with increasing redshift, and that there is 
 a stellar mass where the relation peaks.

Overall the model predicts a $\rho_{\rm SFR,M}$-$M_{\rm stellar}$ relation in reasonable agreement with the observations. 
{The differences between the predictions of the model and the observations are in the range $0.1-0.3$~dex. When integrating over 
stellar mass to give $\rho_{\rm SFR}$, as presented in Fig.~\ref{denstots}, the differences add up to make the large deviations 
shown in Fig.~\ref{denstots}.}
Note that at $z=2$, the Lagos12 model predicts a lower $M_{\rm s}$ than observed. This could be due to the fact that we are not including 
any error source for the stellar mass and SFRs in the model (see \citealt{Marchesini09}; \citealt{Mitchell13}).
We 
show in Fig.~\ref{denstots2} the convolution of our predicted stellar masses with a Gaussian of width 
 $0.2$~dex to illustrate the effect of this uncertainty on the predictions. Once uncertainties are taken into account, 
the predicted $M_{\rm s}$ agrees better with observations.

\subsection{The frequency of highly star-forming galaxies}\label{Sec:SBs}

\begin{figure}
\begin{center}
\includegraphics[trim = 0.5mm 2mm 1mm 1mm,clip,width=0.49\textwidth]{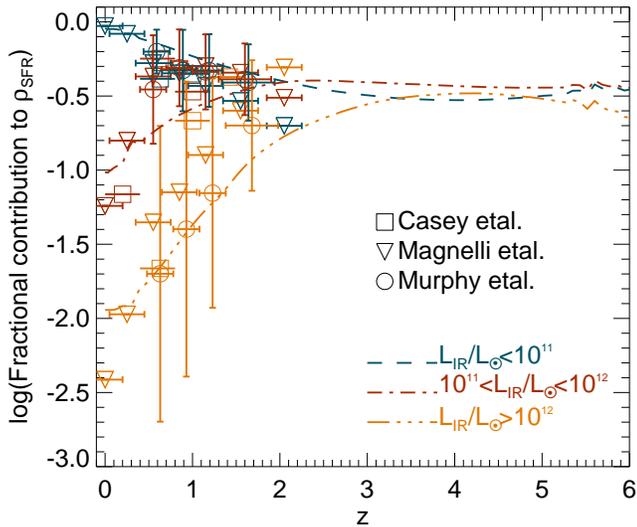}
\caption{Fractional contribution of galaxies with different IR luminosities, as labelled in units of $L_{\odot}$, 
to the global density of SFR for the Lagos12 model, as a function 
of redshift. The observational estimates of \citet{Murphy11}, \citet{Casey12} and \citet{Magnelli13} are shown as symbols. 
For clarity, 
errorbars in the $y$-axis for the observations are shown only for Murphy et al.}
\label{denstotsfrac}
\end{center}
\end{figure}

A commonly used tracer for SF is the IR luminosity, which approximates to the total luminosity emitted by
 interstellar dust, which, in media that are optically thick to UV radiation,
is expected to correlate closely with the SFR. Infrared galaxy surveys have changed our view of the contribution from 
IR luminous galaxies to the IR background, or the global 
SFR density (e.g. \citealt{LeFloch05}). 
From these surveys it is now clear that the contribution from the most luminous IR galaxies to $\rho_{\rm SFR}$ increases 
dramatically with increasing redshift. New mid- and far-IR surveys have allowed the characterisation of the 
 contribution to the IR background and $\rho_{\rm SFR}$ from galaxies with IR luminosities 
down to $L_{\rm IR}\approx 10^{10}L_{\odot}$ between $z=0$ and $z\approx 1.5$. These observations allowed 
the quantification of how more frequent 
bright IR galaxies are at high-redshift compared to the local Universe.
 
We compare in Fig.~\ref{denstotsfrac} the predicted fractional contributions to $\rho_{\rm SFR}$ from galaxies with different IR luminosities 
in the Lagos12 model 
to recent observational estimates from \citet{Murphy11}, which combine measurements at $24\mu$m and $70\mu$m, 
 \citet{Casey12} using Herschel-SPIRE $250\mu$m, $350\mu$m and $500\mu$m bands and \citet{Magnelli13}, which 
use the Herschel-PACS $70\mu$m, $100\mu$m and $160\mu$m bands. The use of several IR bands ensures the detection of the peak 
of the IR spectral energy distribution and an accurate estimate of the total IR luminosity.   
Our model predicts that at $z\gtrsim 3$, the three bins of IR luminosity contribute similarly to $\rho_{\rm SFR}$. 
At lower redshifts the situation is very different. At $z<1$, $\rho_{\rm SFR}$ is dominated by galaxies 
with $L_{\rm IR}< 10^{11}L_{\odot}$, while between $1<z<2$, $\rho_{\rm SFR}$ is dominated by galaxies with $L_{\rm IR}> 10^{11}L_{\odot}$. 
These bright IR galaxies in the Lagos12 model correspond to a combination of normal star-forming galaxies and starbursts. Note that 
the brightest IR galaxies make a negligible contribution to $\rho_{\rm SFR}$ at $z<1.5$, but their contribution increases 
up to $34$\% of the total at $z\approx 3$. We find that the Lagos12  model predicts fractional contributions to $\rho_{\rm SFR}$ that are in good agreement 
with the observations. Although there are systematic differences in the inferred contribution from the brightest 
IR galaxies to $\rho_{\rm SFR}$ between the observational samples, they are still consistent with each other within the errors. 
\citet{Magnelli13} argue that one possible driver of such systematics is the spectroscopic redshift incompleteness, which is different 
for each sample. 

\section{The contribution from star-forming galaxies to the densities of atomic and molecular gas}\label{Contri}
\begin{figure*}
\begin{center}
\includegraphics[trim = 0.5mm 1mm 1mm 0.5mm,clip,width=0.485\textwidth]{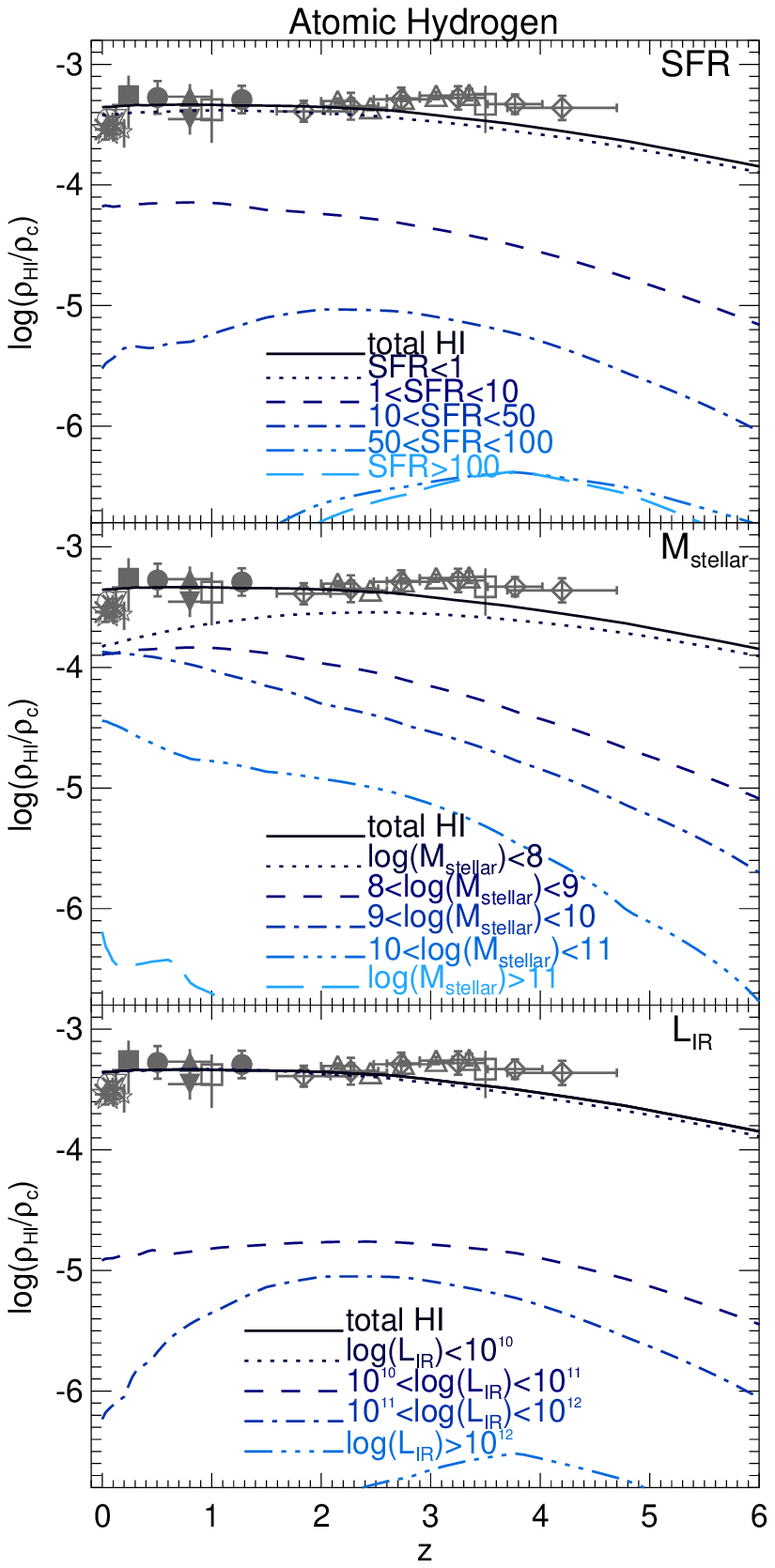}
\includegraphics[trim = 0.5mm 1mm 1mm 0.5mm,clip,width=0.485\textwidth]{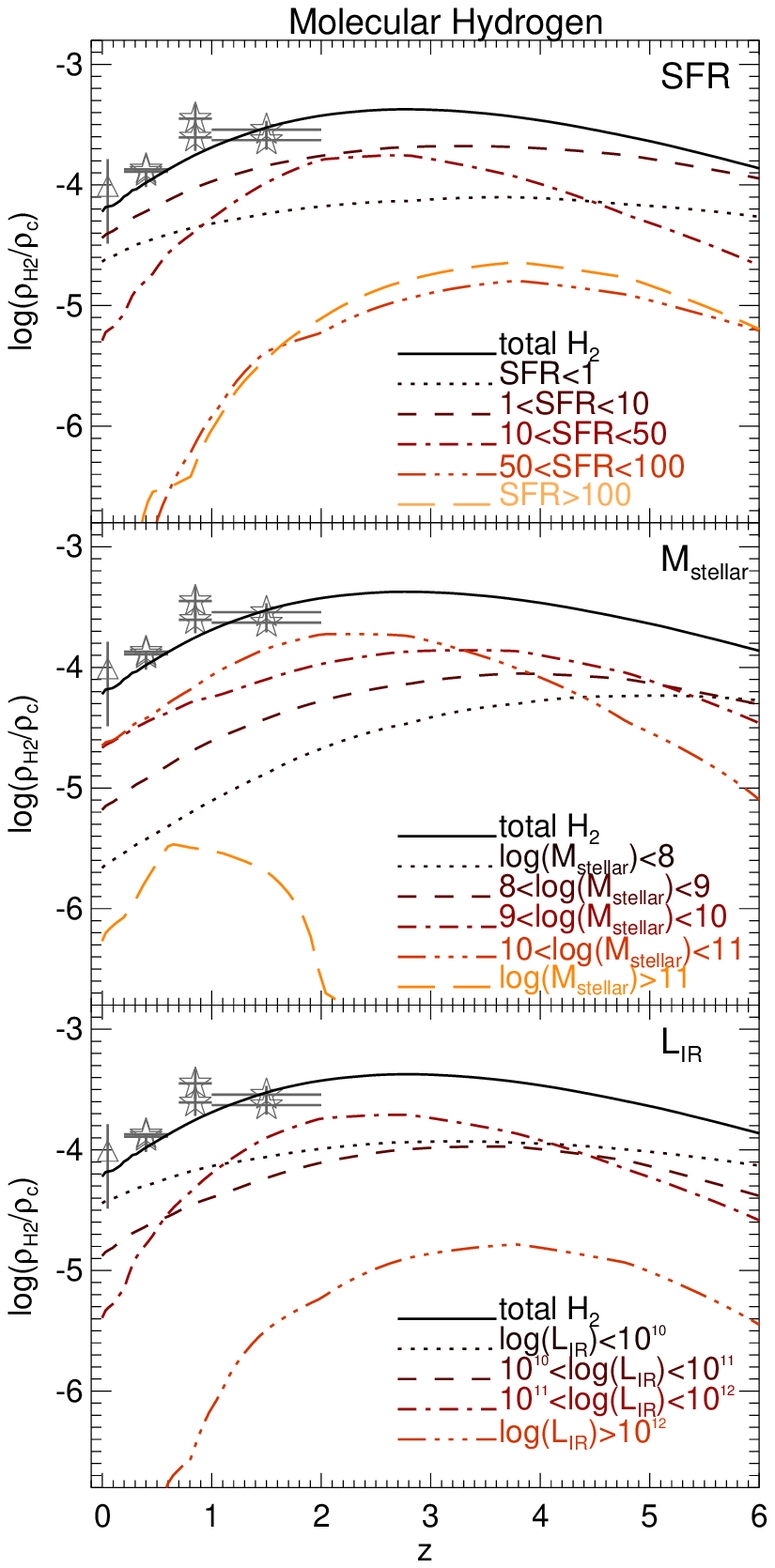}
\caption{Global density of atomic (left panels) and molecular hydrogen (right panels) 
in units of the critical density, as a function of redshift
for the Lagos12 model. The contributions from galaxies with different SFRs (top panels), 
stellar masses (middle panels) and IR luminosities (bottom panels) are shown by 
different lines, as labelled in each panel. The SFRs in the top panels are in units of 
$M_{\odot}\,\rm yr^{-1}$, stellar masses in the middle panels are in $M_{\odot}$ and IR luminosities in the 
bottom panels are in $L_{\odot}$. Observational estimates are as in Fig.~\ref{denstots}.}
\label{denstotsgas}
\end{center}
\end{figure*}
\begin{figure}
\begin{center}
\includegraphics[width=0.5\textwidth]{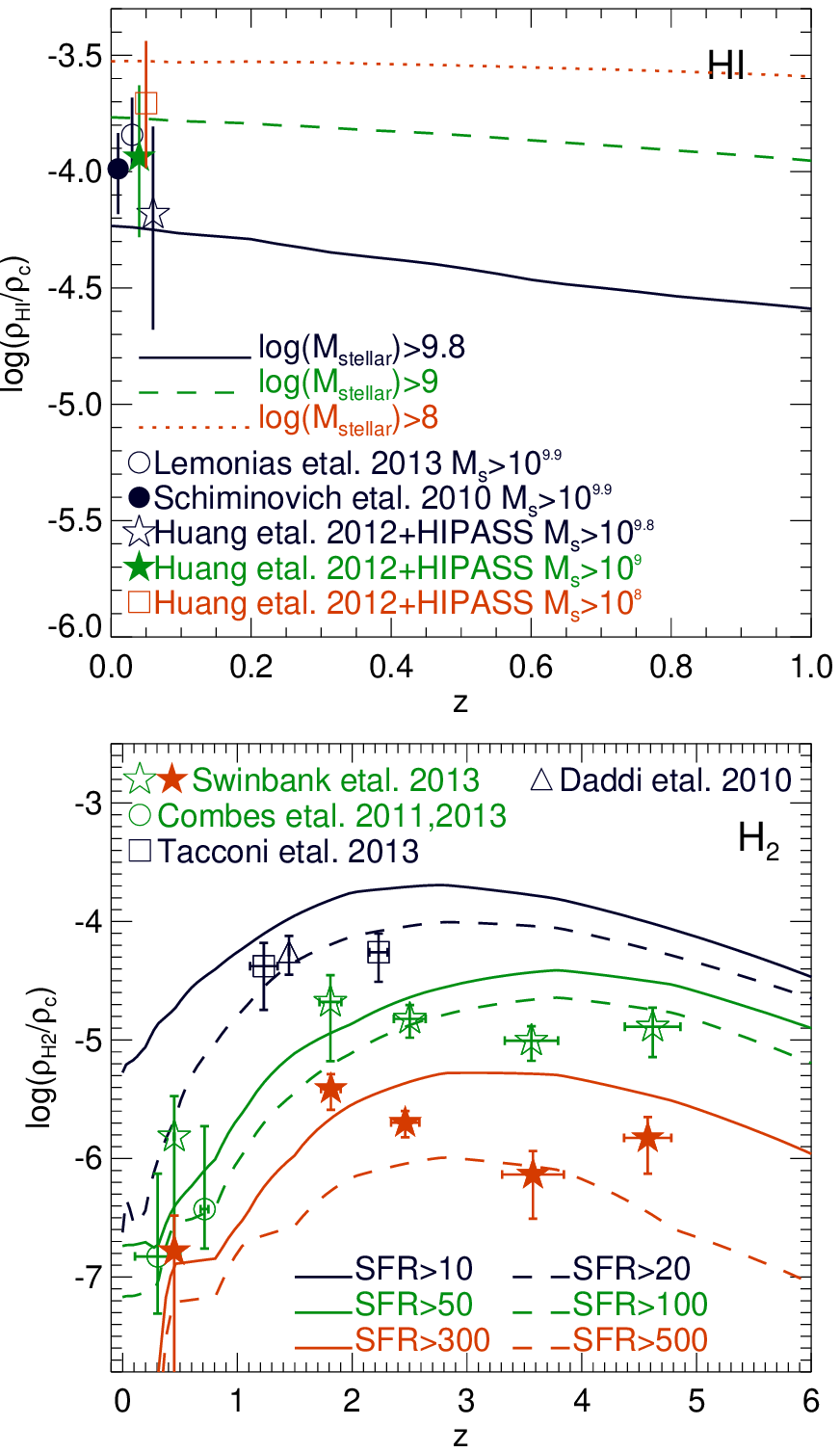}
\caption{{\it Top panel:}  Global density atomic hydrogen 
in units of the critical density, as a function of redshift
for massive galaxies in the Lagos12 model, as labelled. Stellar masses are given in units of $M_{\odot}$. 
We also plot 
the estimate of $\rho_{\rm HI}$ for galaxies with $M_{\rm stellar}>10^{9.9}\,M_{\odot}$ from \citet{Schiminovich10} and 
\citet{Lemonias13}, for three stellar mass limits 
by combining the HI mass function of \citet{Zwaan05} and 
the HI-stellar mass relation of \citet{Huang12}. The stellar mass 
limit in Schiminovich et al. and Lemonias et al. has been corrected from their assumed Chabrier IMF to our adopted 
Kennicutt IMF.
The scatter of the inferred values in the latter estimates is large due to the scatter
in the HI-stellar mass relation.
We thus show three bins of stellar mass 
for model galaxies that should be comparable to the three bins with available observational data.
{\it Bottom panel:} Global density of molecular hydrogen 
in units of the critical density, as a function of redshift
for the most highly star-forming galaxies in the Lagos12 model, as labelled. 
The SFRs are in units of 
$M_{\odot}\,\rm yr^{-1}$. We compare our predicted $\rho_{\rm H_2,SFR}$ with observational inferences from 
\citet{Swinbank13}, which include their sub-millimeter galaxy sample and previous 
samples from \citet{Daddi10}, \citet{Tacconi13} and Combes et al. (2011,2013). The samples of sub-millimeter galaxies from Swinbank et al. 
 is divided into two, those with fluxes at $850\mu$m $>1$~mJy, which roughly corresponds to SFR$\gtrsim 100\,M_{\odot}\rm \,yr^{-1}$, 
 and those with fluxes $>4.2$~mJy, which roughly corresponds to SFR$\gtrsim 300\,M_{\odot}\rm \,yr^{-1}$. 
The samples of Tacconi et al. and Daddi et al. correspond to star-forming galaxies in the main sequence 
of star formation, which have SFR$\gtrsim 20\,M_{\odot}\rm \,yr^{-1}$, while 
 the sample of Combes et al. corresponds to ultra luminous IR galaxies, which roughly have 
SFR$\gtrsim 100\,M_{\odot}\rm \,yr^{-1}$.}
\label{denstotsComp}
\end{center}
\end{figure}

The main goal of this paper is to explore the contribution from galaxies {with different properties} to the overall HI and H$_2$ content  
of the universe. From this we gain an insight into how far observations currently are from tracing the bulk of the 
neutral gas in galaxies in the universe. 
In this section we start by describing the 
contribution from star-forming galaxies to $\rho_{\rm HI}$ and $\rho_{\rm H_{2}}$ and then we analyse the physical 
drivers behind the trends found in $\S$~\ref{Sec:PhysDrivers}. In $\S$~\ref{OtherMods} we investigate how robust these trends are 
by comparing the three {\tt GALFORM} models (described in $\S$~\ref{modelssec}).
We will refer to the HI and H$_2$ densities of galaxies selected by their SFR 
 as $\rho_{\rm HI,SFR}$ and $\rho_{\rm H_{2},SFR}$, respectively, and by stellar mass 
as $\rho_{\rm HI,m}$ and $\rho_{\rm H_{2},m}$, respectively. 
 We remind the reader that both gas components correspond exclusively to gas in galaxies, and that any neutral gas outside galaxies is 
not accounted for. {As in Fig.~\ref{denstots}, we have converted all of the observations to our adopted cosmology ($\S$~\ref{Cosmos}).} 

\subsection{Stellar mass and SFR dependence of $\rho_{\rm HI}$ and $\rho_{\rm H_2}$}

We first discuss the HI content of galaxies, then their H$_2$ 
content and finish with what we expect for galaxies with different IR luminosities.  

{\it HI in galaxies.} The top left-hand panel of 
Fig.~\ref{denstotsgas} shows the evolution of the contribution from galaxies with different SFRs to $\rho_{\rm HI}$
for the Lagos12 model. Most of the galaxies that account for the HI content of the universe have modest SFRs, SFR$<1\,M_{\odot}\rm \,yr^{-1}$, at any time. This 
galaxy population alone makes up $\approx 80$\% of $\rho_{\rm HI}$ 
at $z=0$ and $\approx 94$\% at $z=6$. The remaining HI is found mainly 
in galaxies with SFRs in the range $1<\rm SFR$$/M_{\odot}\,\rm yr^{-1}<10$. Galaxies with high SFRs, $\rm SFR$$>10\,M_{\odot}\,\rm yr^{-1}$, 
only contribute $\approx 0.7$\% to $\rho_{\rm HI}$ at $z=0$ and reach a maximum contribution of $\approx 2.3$\% at $z=3$. The 
form of $\rho_{\rm HI,SFR}$ vs. redshift 
for galaxies with low and high SFRs is very different. The former monotonically decreases with look-back time, 
while the latter have $\rho_{\rm HI,SFR}$ slowly increasing from $z=0$ to $z\approx 2-3$, followed by a gentle decline. 

The middle left-hand panel of 
Fig.~\ref{denstotsgas} shows the evolution of the contribution from galaxies of different stellar masses 
to $\rho_{\rm HI}$ 
for the Lagos12 model. All galaxies, passive and star-forming, have been included in the figure. 
Model galaxies with $M_{\rm stellar}<10^{9}\,M_{\odot}$ dominate the HI content of the universe at $z\lesssim 1$, 
while at $z\gtrsim 1$ most of the HI is locked up in galaxies with $M_{\rm stellar}<10^8\,M_{\odot}$. Our model 
predicts that galaxies with $M_{\rm stellar}>10^{10}\,M_{\odot}$ build up $\approx 14$\% of the HI content of the universe at 
$z=0$. \citet{Schiminovich10} and \citet{Lemonias13} estimated $\rho_{\rm HI,m}$ in the observed 
galaxies of the GALEX Arecibo SDSS Survey (GASS) 
survey, which is a stellar mass selected catalogue of local Universe galaxies (see top panel of Fig.~\ref{denstotsComp}). 
For our adopted IMF, galaxies in the GASS sample have 
 $M_{\rm stellar}>8.9\times 10^{9}\,M_{\odot}$. We calculate $\rho_{\rm HI,m}$ of model galaxies selected using the same 
stellar mass cut at $z=0$ and find that our prediction is a factor $\approx 0.1-0.2$~dex lower than the observational inference. 
The exact factor depends on the treatment of the non-detections in the observations. 
This difference could be partially explained 
 as being driven by cosmic variance; our simulated volume is relatively small, and we lack the very gas-rich galaxies with HI masses 
comparable to the most gas-rich galaxies included in Schiminovich et al. and Lemonias et al. 
to infer $\rho_{\rm HI,m}$ of massive galaxies. Another source of discrepancy is related to the crude treatment of 
stripping of HI gas in galaxy groups and clusters. In our model, when galaxies become satellites, they instantaneously 
lose their hot gas reservoir (i.e. strangulation of hot gas), transferring it to the central halo 
while slowly consuming ther remaining gas in the ISM. 
A more accurate treatment may be needed to explain the neutral gas content of galaxies in groups and clusters (Lagos, Davis et al. 
in preparation). 
 
With the aim of making a more quantitative comparison to lower stellar mass bins, we calculate $\rho_{\rm HI,m}$ 
for galaxies with $M_{\rm stellar}>10^8\,M_{\odot}$, $M_{\rm stellar}>10^9\,M_{\odot}$ and $M_{\rm stellar}>8.9\times 10^{9}\,M_{\odot}$ 
by combining the observed HI mass function of \citet{Zwaan05} with the HI-stellar mass relation found by \citet{Huang12} 
for the ALFALFA sample, corrected to a Kennicutt IMF. 
The Lagos12 model predictions show good agreement with these observationally inferred 
 $\rho_{\rm HI,m}$ in the three mass bins. The scatter of the observationally inferred $\rho_{\rm HI,m}$ is rather large 
due to the scatter in the HI-stellar mass relation.  

Compared to previous theoretical estimates, we find that our predictions for $\rho_{\rm HI,m}$ agree well with the predicted trends 
in \citet{Dave13}. Dav\'e et al. find that most of HI in the universe, which in their case corresponds to both HI in galaxies and 
in the intergalactic medium, is locked up in galaxies with stellar masses $\lesssim 10^8\,M_{\odot}$ at 
$z>1.5$. At $z<1.5$, Dav\'e et al. find that galaxies with stellar masses in the range $10^8-10^9\,M_{\odot}$ 
 increase their contribution to $\rho_{\rm HI}$ to similar values as the lower stellar mass galaxies. These trends 
are similar to the ones we find for the Lagos12 model. \citet{Popping13} include in a semi-analytic model of galaxy formation 
similar SF laws to those that 
were developed in \citet{Lagos10} and found that the HI density is dominated by galaxies with $M_{\rm stellar}<10^7,M_{\odot}$, 
which is even lower than in our case. These predictions from Popping et al. may be in slight tension with the observationally inferred 
contribution from galaxies with $M_{\rm stellar}>10^8,M_{\odot}$ shown in Fig.~\ref{denstotsgas}, 
which can account for $\approx 70$\% of the observed HI mass density at $z=0$, albeit with large errorbars.  

\begin{figure}
\begin{center}
\includegraphics[trim = 1.5mm 7.5mm 1mm 1mm,clip,width=0.45\textwidth]{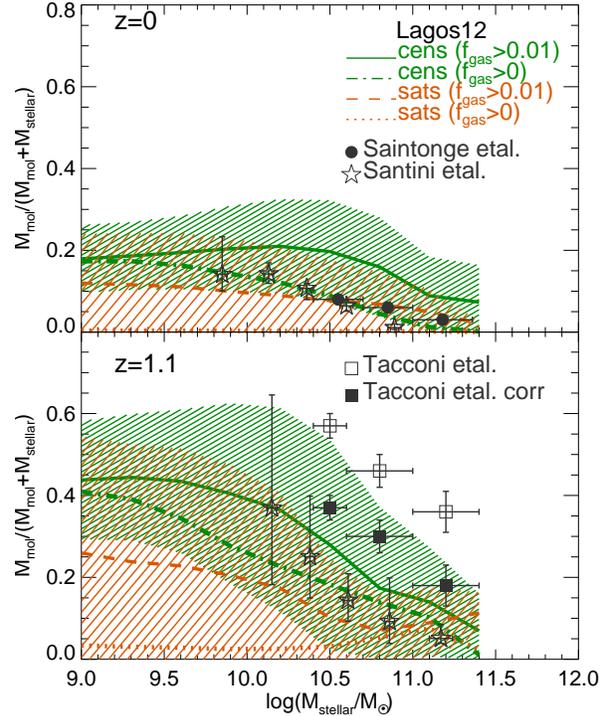}
\caption{Molecular gas fraction, defined as $f_{\rm gas}=M_{\rm mol}/(M_{\rm mol}+M_{\rm stellar})$, where $M_{\rm mol}$ 
includes hydrogen and helium, as a function of 
stellar mass, at $z=0$ (top panel) and $z=1.1$ (bottom panel) for the Lagos12 model. 
Central and satellite galaxies are shown 
separately, each for two selections in molecular gas fraction, $f_{\rm gas}>0$ and $f_{\rm gas}>0.01$. 
The aim of this is to show
how sensitive the predicted trends are to selection effects that are typically present in observations. 
Lines show the median of the relations, and the shaded regions 
represent the $10$ and $90$ percentiles and are shown only for the case of centrals and satellites with $f_{\rm gas}>0.01$. 
At $z\approx 0$, filled circles 
show the observationally inferred trend from the carbon monoxide survey of \citet{Saintonge11}, 
at $z\approx 0$ and $z=1.1$ stars {show H$_2$ inferred masses from the dust measurements 
of \citet{Santini13} using Herschel}, 
 and the $z=1.1$ observations of \citet{Tacconi13} of carbon monoxide in normal star-forming galaxies 
with and without 
the incompleteness correction are shown as filled and empty squares, respectively. 
{Note that we express masses in physical units. For these we corrected all of the observations 
to our adopted cosmology ($\S$~\ref{Cosmos}).}}
\label{denstotsComp2}
\end{center}
\end{figure}

{\it H$_2$ in galaxies.} The behaviour of star-forming galaxies in the $\rho_{\rm H_2}$-z plane, shown in the top and middle right-hand panels 
of Fig.~\ref{denstotsgas}, contrasts with that in the 
$\rho_{\rm HI}$-z plane. Galaxies with SFR$<1\,M_{\odot}\rm \,yr^{-1}$ represent $\approx 35$\% of the H$_2$ at $z=0$ and 
only $\approx 15$\% at $z=3$. Galaxies with $1<\rm SFR$$/M_{\odot}\,\rm yr^{-1}<10$ contribute $\approx 56$\% of the H$_2$ at $z=0$, and 
their contribution also decreases with increasing redshift, yielding $\approx 40$\% at $z=2$. One of 
the most interesting predictions of the top right-hand panel 
of Fig.~\ref{denstotsgas} is that galaxies with $10<\rm SFR$$/M_{\odot}\,\rm yr^{-1}<50$ contain a large fraction of the $H_2$ of 
the universe at high-redshift, reaching a maximum contribution of $\approx 40$\% at $z=2$. Highly star-forming galaxies are also 
more important in H$_2$ than HI, reaching a maximum contribution of $\approx 9$\% in the redshift range $z\approx 3.5-4.5$. As 
in $\rho_{\rm HI}$, the functional form of $\rho_{\rm H_{2},SFR}$ of highly star-forming galaxies is very different from that 
of galaxies with modest SFRs. Galaxies with SFR$<10\,M_{\odot}\rm \,yr^{-1}$ give a $\rho_{\rm H_{2},SFR}$ slowly increasing with redshift 
up to $\approx 3.5$, followed by a very slow decline at higher redshifts, 
while galaxies with SFR$>10\,M_{\odot}\rm \,yr^{-1}$ show $\rho_{\rm H_{2},SFR}$
 increasing by more than a factor $\approx 30$ between $z=0$ and $z\approx 2$. 

Observational estimates of $\rho_{\rm H_2,SFR}$ are very scarce and subject to strong systematics, 
such as the choice of CO-H$_2$ conversion factor, the dust-to-gas mass ratio and the characterisation of the selection function.
Even though these uncertainties limit current possibilities to derive the quantities above, it is possible to use 
available CO surveys of highly star-forming galaxies to infer $\rho_{\rm H_2,SFR}$ to within a factor of $\approx 2-3$. 
\citet{Swinbank13} infer $\rho_{\rm H_2,SFR}$ for galaxies with SFR$\gtrsim 30\,M_{\odot}\rm \,yr^{-1}$.
Swinbank et al. analyse the H$_2$ abundance of 
sub-millimeter galaxies, together with three other 
samples of galaxies with SFRs~$\gtrsim 100\,M_{\odot}\rm \,yr^{-1}$ from 
\citet{Combes11} and \citet{Combes13}, and with SFRs~$\gtrsim 20\,M_{\odot}\rm \,yr^{-1}$ 
from \citet{Daddi10} and \citet{Tacconi13}. To calculate $\rho_{\rm H_2}$ from the samples of Tacconi et al. and 
Combes et al. is not straightforward, given the complexity of the selection functions. With this in mind, Swinbank et al. 
 used simulated galaxy catalogues from the Millennium database to select model galaxies 
using the same observational selection criteria (in H$_{\alpha}$ or IR luminosity, optical luminosity and stellar mass), 
 which informed the number density of galaxies that resemble the samples of Tacconi et al. and Combes et al..
Swinbank et al. used these number densities to estimate $\rho_{\rm H_2}$ from the samples of Tacconi et al. and Combes et al.

Note that the SFR lower limit in the observed samples is approximate and can vary by a factor of $\approx 2$. We compare the estimates 
of Swinbank et al. with our predicted $\rho_{\rm H_2,SFR}$ for these highly star-forming galaxies in the bottom panel 
of Fig.~\ref{denstotsComp}. 
 To illustrate the effect of varying 
the SFR limits by a factor of $2$, we show in Fig.~\ref{denstotsComp} $\rho_{\rm H_2,SFR}$ for galaxies with 
SFR$\gtrsim 50\,M_{\odot}\,\rm yr^{-1}$ and SFR$\gtrsim 100\,M_{\odot}\,\rm yr^{-1}$, which are compared to the 
inferences of $\rho_{\rm H_2,SFR}$ by Swinbank et al. for the samples of 
Combes et al. and their faint sub-millimeter galaxies (SMGs), 
and with SFR$\gtrsim 300\,M_{\odot}\,\rm yr^{-1}$ and SFR$\gtrsim 500\,M_{\odot}\,\rm yr^{-1}$, which are compared to the 
sample of Swinbank et al. of bright SMGs.
 To compare with the estimates presented by Swinbank et al. using the Daddi et al. and Tacconi et al. samples, 
we show the predicted $\rho_{\rm H_2,SFR}$ for galaxies with 
SFR$\gtrsim 10\,M_{\odot}\,\rm yr^{-1}$ and SFR$\gtrsim 20\,M_{\odot}\,\rm yr^{-1}$.
 Swinbank et al. argue that there are systematic effects in the 
estimates of the number densities of sources in addition to other systematics effects related to the 
CO-to-H$_2$ conversion factor. This implies that the errorbars 
on the observational data in Fig.~\ref{denstotsComp} are likely to be lower limits to the true uncertainty. 
 Note that the samples of Daddi et al., Tacconi et al. and Combes et al. correspond to observations of CO, while 
Swinbank et al. used inferred dust masses from their FIR observations to estimate  H$_2$ 
 masses by adopting a dust-to-gas mass ratio.  
The model predictions are in good agreement with the observations, within the errorbars, over the whole redshift range.  
{Note that the data here gives independent constrains on our model than those provided by Berta et al. and shown in 
Fig.~\ref{denstots}.}
%Unlike the datapoints from Berta et al. in Fig.~\ref{denstots}, the estimates of $\rho_{\rm H_2,SFR}$ in Fig.~\ref{denstotsComp} do not 
%depend on SFR, and therefore provide an additional, independent test to our model predictions. 
Galaxies 
 used to calculate $\rho_{\rm H_2,SFR}$ in the observations, as in our model predictions, 
are a mixture of merging systems and normal star-forming galaxies 
(which are referred to in our model as quiescent galaxies), which has been concluded 
from the measured velocity fields of the CO emission lines 
in some of the observed galaxies.

The trends between $\rho_{\rm H_{2},m}$ and stellar mass in the middle right-hand panel of
Fig.~\ref{denstotsgas} are again very different from those obtained for $\rho_{\rm HI}$. 
We find that $\rho_{\rm H_{2}}$ is dominated by relatively massive galaxies, 
$10^{10}\,M_{\odot}<M_{\rm stellar}<10^{11}\,M_{\odot}$, from $z=0$ up to $z\approx 3$. 
Note that the contribution from these galaxies to $\rho_{\rm H_{2}}$ increases from $42$\% at $z=0$ to $52$\% at $z=1.5$. 
At $z\gtrsim 3$, galaxies with lower stellar masses, $10^{8}\,M_{\odot}<M_{\rm stellar}<10^{10}\,M_{\odot}$, 
dominate the H$_2$ content. It is also interesting to note that \citet{Karim11} and \citet{Sobral13} find that the SFR density 
 is dominated by galaxies with stellar masses in the range 
$1.4\times 10^{10}\,M_{\odot}<M_{\rm stellar}<10^{11}\,M_{\odot}$, which overlaps with the stellar mass range we find to dominate 
 $\rho_{\rm H_2}$. {Note that the contribution from galaxies with $M_{\rm stellar}<10^8\,M_{\odot}$ is resolved, as higher resolution 
runs produce the same small contribution agreeing to a factor of better than $5$\% (see Appendix~\ref{Resolution}).}

In Fig.~\ref{denstotsComp2}, we compare the relation between the molecular gas fraction and the stellar mass in the model 
with observations {of CO emission lines (\citealt{Saintonge11}; \citealt{Tacconi13}) and 
dust masses \citep{Santini13}}. From this figure we can draw two main conclusions. First, 
our model predicts a strong evolution of the molecular gas fraction that agrees well with the observationally inferred one 
(see also \citealt{Geach11} for a former comparison), and second, the comparison with observations needs to be done carefully, 
as the predicted evolution strongly depends on galaxy properties, such as the gas fraction itself, stellar mass or halo mass. 
Observations other than at $z=0$ are strongly biased towards high molecular gas fractions, and therefore observations at 
 $z=1.1$ should be compared with the evolution predicted for galaxies with molecular gas fractions that are above $1$\%. 
Another interesting aspect is the difference between central and satellite galaxies, which comes mainly from the lack of cooled 
gas accretion in satellites. \citet{Lagos10} show that most of the galaxies in the main sequence of the SFR-$M_{\rm stellar}$ plane 
correspond to central galaxies in the model, and therefore observed galaxies, which also fall in this main sequence,  
should be compared to our expectations for centrals.  

{\it The dependence on IR luminosity.} In the bottom panels of Fig.~\ref{denstotsgas} we show the contribution to the HI and H$_2$ densities 
from galaxies with different IR luminosities. This quantity is more easily inferred from observations than the SFR 
or stellar masses, which, 
as we discussed in $\S$~\ref{cosmicevo}, are affected by systematic uncertainties that are not accurately constrained. 
The IR luminosity, on the other hand, is well estimated from photometry in few IR bands (e.g. \citealt{Elbaz10}). 
The trends seen with IR luminosity in Fig.~\ref{denstotsgas} 
are similar to those obtained for the SFR (top panels of Fig.~\ref{denstotsgas}). 
This is due to the good correlation between the IR luminosity and the SFR predicted by the model. However, the normalisation 
of the $L_{\rm IR}$-SFR relation 
is predicted to vary with SFR and redshift, due to the underlying evolution of gas metallicity and the gas density in the 
ISM of galaxies. 

\subsection{The physical drivers of the gas content of star-forming galaxies}\label{Sec:PhysDrivers}

\begin{figure}
\begin{center}
\includegraphics[width=0.5\textwidth]{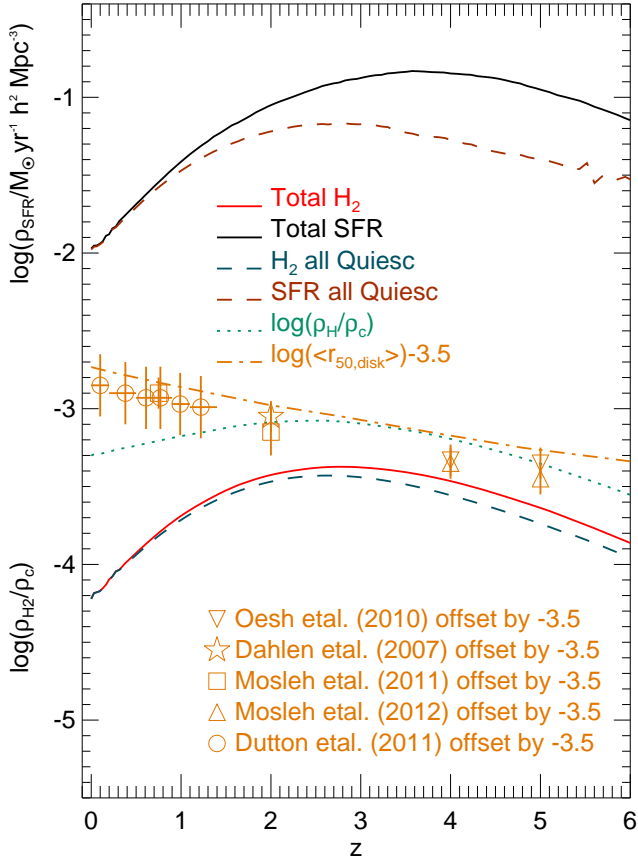}
\caption{Global density of the SFR in units of $M_{\odot}\, {\rm yr}^{-1}\, h^2\,{\rm Mpc}^{-3}$ (black line), and of the 
 molecular hydrogen (red line) in units of the critical density,
as a function of redshift for the Lagos12 model. We show the contribution from quiescent star formation (i.e. taking place in 
the galaxy disk) as dashed lines, the evolution of the neutral hydrogen gas, atomic plus molecular, in the ISM of galaxies 
in units of $\rho_{\rm c}$ as dotted line and the evolution of the median half-mass radius,$\langle r_{\rm disk}\rangle$, of late-type 
galaxies with stellar masses in the range $7\times 10^{9}\,M_{\odot}<M_{\rm stellar}<10^{10}\,M_{\odot}$, 
as dot-dashed line. The latter range is chosen to make it comparable to the observational inferences.  
The units of $\langle r_{\rm disk}\rangle$ are $\rm kpc$, and 
$\rm log($$\langle r_{\rm disk}\rangle)$ are plotted offset by $-3.5$ to fit in the figure. 
 We also show the observational estimate of the half-light radius evolution {from \citet{Dutton11} 
\citet{Dahlen07}, \citet{Oesch10}, \citet{Mosleh11} and \citet{Mosleh12} for galaxies with stellar masses $\sim 8.9\times 10^{9}\,M_{\odot}$.
 Note that the errorbars in the data of Dutton et al. show the $1\sigma$ dispersion around the median of the size-mass relation, 
while the other observations quote Poisson errors and are therefore not necessarily representative of the true dispersion of the 
size-mass relation. The units of the observational measurements of $r_{\rm disk}$ are kpc and are 
shifted by $-3.5$ dex, as are the model predictions.} 
{The units for densities are comoving, as in Fig.~\ref{denstots}. In the case of the half-light radius, the units are physical and 
therefore we have scaled the observations to our adopted cosmology ($\S$~\ref{Cosmos}).}}
\label{denstotsQuiesc}
\end{center}
\end{figure}

\begin{figure}
\begin{center}
\includegraphics[width=0.5\textwidth]{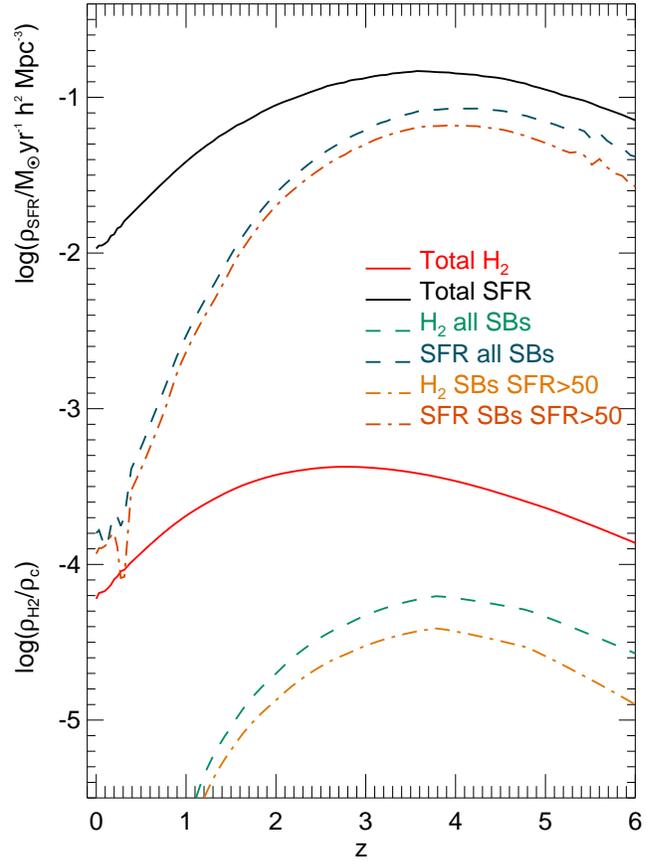}
\caption{Global density of the SFR in units of $M_{\odot}\, {\rm yr}^{-1}\, h^2\,{\rm Mpc}^{-3}$ (black line), and of the 
 molecular hydrogen (red line) in units of the critical density,
as a function of redshift for the Lagos12 model. We show the contribution from all starburst galaxies (i.e. those with 
at least 50\% of their global SFR taking place in a starburst) to $\rho_{\rm SFR}$ and $\rho_{\rm H_{2}}$ 
as dashed lines, and from starburst galaxies with 
SFR~$\ge 50~M_{\odot}\, \rm yr^{-1}$ as dot-dashed lines. {The units for densities are comoving, as in Fig.~\ref{denstots}.}}
\label{denstotsSBs}
\end{center}
\end{figure}

There are three key physical processes that drive the trends with SFR and stellar mass in Fig.~\ref{denstotsgas}; 
 (i) the regulation between gas outflows and accretion of gas onto galaxies, (ii) the size evolution of normal 
star-forming galaxies (i.e. which undergo quiescent SF) and its effect on the gas surface density, and (iii) 
the frequency of starbursts and its evolution. 
Below we summarise the main effects each of these physical processes have on $\rho_{\rm SFR}$, $\rho_{\rm HI}$ and $\rho_{\rm H_2}$.

{\it Evolution in $\rho_{\rm cold}$.} The evolution of the total density of 
cold gas $\rho_{\rm cold}=\rho_{\rm HI}+\rho_{\rm H_2}$ is shown in Fig.~\ref{denstotsQuiesc}. Also shown are 
the evolution of $\rho_{\rm SFR}$ and $\rho_{\rm H_2}$ 
for all galaxies, for all quiescent galaxies, 
and the median half-mass radius of all galaxies with 
stellar masses in the range $7\times 10^{9}\,M_{\odot}<M_{\rm stellar}< 10^{10}\,M_{\odot}$. 

Quiescent SF dominates the SFR in the universe at $z\lesssim 2.8$, and 
at higher redshift it contributes $\approx 30-50$\% to $\rho_{\rm SFR}$. 
 The peak of $\rho_{\rm SFR}$ for quiescent galaxies takes place at lower 
redshift than for starbursts. 
The evolution of $\rho_{\rm SFR}$ of quiescent galaxies is very similar to the 
evolution of $\rho_{\rm H_2}$, 
reassuring our previous statement that starbursts contribute very little to $\rho_{\rm H_2}$. 
The latter closely follows $\rho_{\rm cold}$ at $z\gtrsim 2$. 
The evolution of $\rho_{\rm cold}$ is driven by the balance between 
accretion of gas and outflows from galaxies. On average, accretion and outflows in galaxies at a given time 
self-regulate in a way that the SFR-$M_{\rm stellar}$ relation is characterised by a small dispersion and a power-law 
index of $\approx 0.8-0.9$ \citep{Lagos10}. 
Small perturbations to this self-regulation phase 
of galaxy growth drive the slow evolution of $\rho_{\rm cold}$ and of the normalisation of the 
SFR-$M_{\rm stellar}$ relation of normal star-forming galaxies. 
In the regime where $\rho_{\rm HI}\gg \rho_{\rm H_2}$, which is the case at $z\lesssim 1$, $\rho_{\rm HI}$ follows very closely the 
evolution of $\rho_{\rm cold}$.
 At $z\gtrsim 4$, the ratio between $\rho_{\rm H_2}$ and $\rho_{\rm HI}$  
evolves slowly, and therefore $\rho_{\rm H_2}$ and $\rho_{\rm HI}$ decrease 
with increasing redshift at a similar rate as $\rho_{\rm cold}$ decreases. This can be seen
in the evolution of the former two quantities shown in Fig.~\ref{denstots} and 
of $\rho_{\rm cold}$ shown in Fig.~\ref{denstotsQuiesc}. In the intermediate redshift regime, 
$2<z<4$,  
$\rho_{\rm H_2}\gtrsim \rho_{\rm HI}$, and therefore $\rho_{\rm H_2}$ peaks at a similar redshift 
as $\rho_{\rm cold}$. 

At $z<3$, the evolution of $\rho_{\rm H_2}$ and $\rho_{\rm SFR}$ of quiescent galaxies is more dramatic than 
it is for $\rho_{\rm cold}$. This is due to the fact that the ratio 
$\rho_{\rm H_2}/\rho_{\rm HI}$ for quiescent galaxies is strongly evolving 
 due to the change in the surface density of gas plus stars, which change the 
hydrostatic pressure in the midplane of disks. 
 A factor that strongly affects the evolution of the surface density of gas and stars
 is the evolution in the sizes of galaxies.

{\it Evolution in size.} On average, galaxy sizes increase with time (see Fig.~\ref{denstotsQuiesc}).
What drives the growth in the sizes of galaxy disks with time is 
 the angular momentum of the accreted gas. As the universe expands, 
gas that gets to galaxies comes from further out, which means that it brings higher angular momentum 
compared to gas that was accreted in the past. If the angular momentum loss 
is small, the newly cooled gas that is accreted settles down at an outer radius, increasing the half-mass radius (see \citealt{Cole00}
 for a quantitative description of the size evolution of galaxies in {\tt GALFORM}).
{We compare the evolution of sizes in the model galaxies with the observed increase reported.
Note that in the case of \citet{Dahlen07}, \citet{Oesch10} and Mosleh et al. (2011,2012) there is no morphological selection 
and the size evolution is measured for all galaxies, while in the case of \citet{Dutton11}, the reported 
half-light radius evolution is for late-type galaxies only. In all cases we plot the reported evolution 
for galaxies with stellar masses $M_{\rm stellar}\sim 8.9\times 10^{9}\,M_{\odot}$.}
The stellar masses of the observed samples have been scaled from a 
\citet{Chabrier03} to a \citet{Kennicutt83} IMF.
{In the model we select} all late-type galaxies (i.e. bulge-to-total stellar mass ratio $<0.5$) 
 with stellar masses in the range 
$7\times 10^{9}\,M_{\odot}<M_{\rm stellar}<10^{10}\,M_{\odot}$ and calculate the median half-mass radius of their disks. 
We find that the growth rate of galaxies agrees well with the observations, {although with small discrepancies at higher redshifts 
towards larger radii. The observed samples at these high-redshifts are very small and encompass all galaxies, late- and early-type, which 
can bias the size-stellar mass relation toward smaller sizes. In addition, only Poisson errors have been reported which makes it  
difficult to estimate the significance of the deviations above. At $z>5$ there are only a handful of galaxies with stellar masses in the range 
we plot and therefore we do not show them in this figure.}
 
The size evolution has an effect on the 
gas surface density, and therefore on $\rho_{\rm H_2}/\rho_{\rm HI}$ and $\rho_{\rm SFR}$.
Quantitatively, $\rho_{\rm H_2}$ decreases by $\approx -0.9$~dex from $z\approx 2.5$ to $z=0$. In the same redshift range, 
$\rho_{\rm cold}$ decreases by $\approx -0.3$~dex and the median size of late-type galaxies in the model 
increases by $0.4$~dex. From combining the evolution in $\rho_{\rm cold}$ and $\langle r_{\rm 50} \rangle$ 
in the regime where $\Sigma_{\rm cold}\gg \Sigma_{\rm stellar}$, we can derive the expected 
change in $\rho_{\rm H_2}$ to be $-1.2$~dex. This number is an upper limit on the magnitude of the variations 
in $\rho_{\rm H_2}$ between $\approx 2.5$ and $z=0$, as the contribution from stars to the hydrostatic pressure, on average, 
increases with time (see L11). 

{\it The frequency of starbursts.} Fig.~\ref{denstotsSBs} shows $\rho_{\rm SFR}$ and $\rho_{\rm H_2}$ for all galaxies, 
for all starburst galaxies, and starbursts with $\rm SFR$$>50\,M_{\odot}\,\rm yr^{-1}$.
Starbursts make an important contribution to the global SFR at $z\gtrsim 3$. However, their contribution to 
$\rho_{\rm H_2}$ is less important. The reason for this is that the SF law adopted for starbursts in our model 
is typically more efficient at converting gas into stars than that adopted for quiescent SF (see $\S$~\ref{modelssec}). 
Starbursts with SFR$>50\,M_{\odot}\,\rm yr^{-1}$ contribute about half of the SFR and H$_2$ mass density coming from 
all starbursts. The contribution from these high SFR starbursts to $\rho_{\rm H_2,SFR}$ of all  
galaxies, quiescent and starbursts, with $\rm SFR$$>50\,M_{\odot}\,\rm yr^{-1}$ is large, and this can be seen from the 
similar shape and normalisation of the curve for starbursts with  $\rm SFR$$>50\,M_{\odot}\,\rm yr^{-1}$ in  Fig.~\ref{denstotsSBs} 
and all galaxies with  $\rm SFR$$>50\,M_{\odot}\,\rm yr^{-1}$ in Fig.~\ref{denstotsgas}.

In fact, the $\rho_{\rm H_2,SFR}$ from all galaxies with $\rm SFR$$>50\,M_{\odot}\,\rm yr^{-1}$ and $\rm SFR$$>100\,M_{\odot}\,\rm yr^{-1}$  
shown in the top
right-hand panel of Fig.~\ref{denstotsgas}, increases by $\approx 2-2.3$~dex from $z=0$ to $z\approx 3.5$.
This increase is slightly larger than that of $\approx 2$~dex in 
$\rho_{\rm SFR}$ and $\rho_{\rm H_2,SFR}$ of starburst galaxies with $\rm SFR$$>50\,M_{\odot}\,\rm yr^{-1}$ shown 
in Fig.~\ref{denstotsSBs}. 
The small difference between the two is due to the contribution from 
quiescent galaxies with $\rm SFR$$>50\,M_{\odot}\,\rm yr^{-1}$, which also  
have $\rho_{\rm H_2,SFR}$ increasing with increasing redshift, but more slowly 
compared to starbursts. 
Note that the strong increase of $\rho_{\rm SFR}$ and $\rho_{\rm H_2}$ of starbursts 
is also largely responsible for the evolution of galaxies with $L_{\rm IR}>10^{12}\,L_{\odot}$ in Fig.~\ref{denstotsfrac}. 
The contribution from starbursts to $\rho_{\rm HI}$ is negligible. 

\subsection{The HI and H$_2$ gas contents of galaxies in the 
Lacey14 and Gonzalez-Perez14 models}\label{OtherMods}

\begin{figure}
\begin{center}
\includegraphics[trim = 1.5mm 1.5mm 1mm 1mm,clip,width=0.47\textwidth]{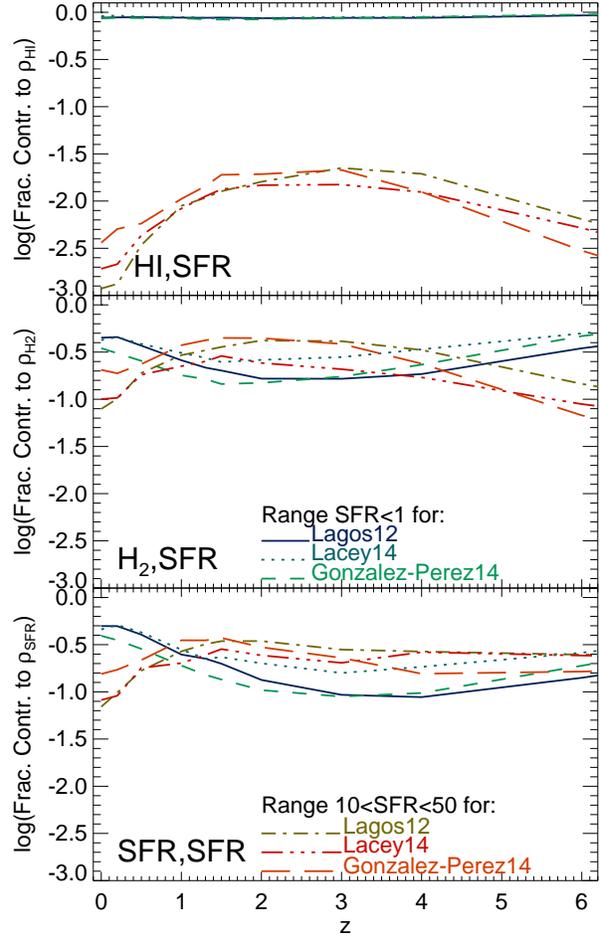}
\caption{Fractional contribution to the global density of HI (top panel), 
H$_2$ (middle panel) and SFR (bottom panel) 
from galaxies in two bins of SFR, SFR$<0.1\,M_{\odot}\,\rm yr^{-1}$, as labelled in the middle panel, and 
$10\,M_{\odot}\,{\rm yr}^{-1} <{\rm SFR}<50\,M_{\odot}\,\rm yr^{-1}$, as labelled in the bottom panel, for 
the Lagos12, Lacey14 and Gonzalez-Perez14 models, 
 as a function of redshift.}
\label{FracSFRs}
\end{center}
\end{figure}

\begin{figure}
\begin{center}
\includegraphics[trim = 1.5mm 1.5mm 1mm 1mm,clip,width=0.47\textwidth]{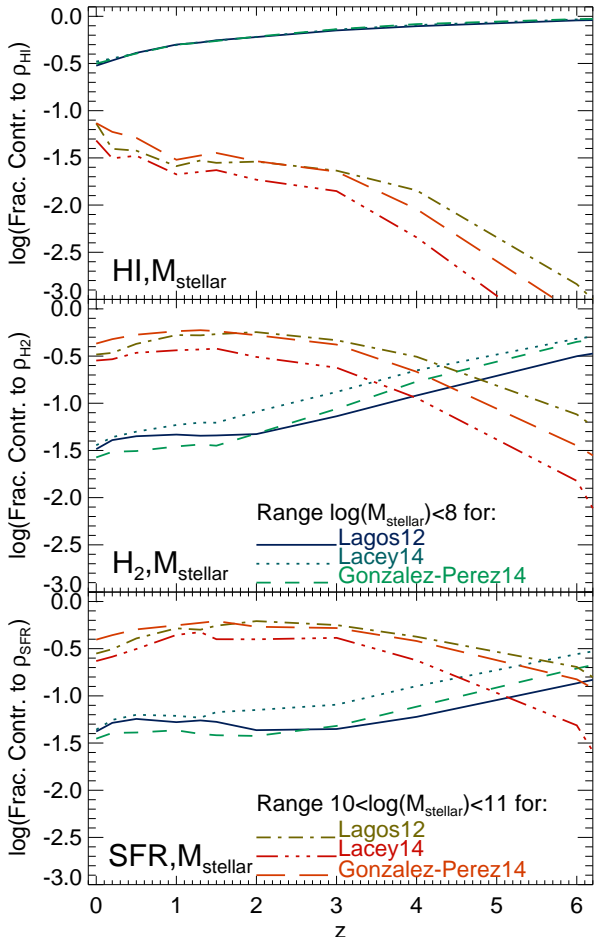}
\caption{As in Fig.~\ref{FracSFRs}, but for galaxies 
 in two bins of stellar mass, $M_{\rm stellar}<10^8\,M_{\odot}$, as labelled in the middle panel, and 
$10^{10}\,M_{\odot}<M_{\rm stellar}<10^{11}\,M_{\odot}$, as labelled in the bottom panel.}
\label{FracMs}
\end{center}
\end{figure}

One of the aims we have when analysing different {\tt GALFORM} models 
is to get a better insight into how much variation we 
expect in the predicted contributions from galaxies selected according to their 
stellar mass or SFR to the densities of HI and H$_2$. This also helps to illustrate how 
some physical processes, other than the SF law, affect the predictions 
described in $\S$~\ref{Contri}. From this, it is also possible to identify how robust 
the trends described in $\S$~\ref{Contri} are.

Fig.~\ref{FracSFRs} shows the fractional contribution from galaxies in two different bins of SFR to the densities of 
HI, H$_2$ and SFR in the Lagos12, Lacey14 and Gonzalez-Perez14 models. {The bins were chosen to match 
those of Fig.~\ref{denstotsgas}}. 
We regard the following trends as 
robust as they are present in all three models: (i) galaxies with low SFRs dominate 
$\rho_{\rm HI}$, while for $\rho_{\rm H_2}$ and $\rho_{\rm SFR}$, they make a modest contribution that decreases from 
$z=0$ to $z\approx 2.5$, (ii) galaxies with high SFRs have a contribution to $\rho_{\rm HI}$ that increases with increasing 
redshift, at least up to $z\approx 2.5$, but that never reach more than few percent, 
for $\rho_{\rm H_2}$. Their contribution is of $\approx 8-15$\% at $z\sim 0$, increasing 
to $\approx 25-40$\% at $z\approx 1-2.5$, and for $\rho_{\rm SFR}$ their contribution increases with redshift 
up to $z\approx 1.5$ and stays relatively constant at higher redshifts. 
Note that the exact peak and normalisation of the contributions of each 
model sample to the densities of HI, H$_2$ and SFR depend on the model. 

Fig.~\ref{FracMs} shows the fractional contribution from galaxies in two different bins of stellar mass to 
$\rho_{\rm HI}$, $\rho_{\rm H_2}$ and $\rho_{\rm SFR}$ in the Lagos12, Lacey14 and Gonzalez-Perez14 models.
{The bins were chosen to match                    
those of Fig.~\ref{denstotsgas}.} 
 Here, the following trends are robust: (i) $\rho_{\rm HI}$ has always a large contribution from 
low mass galaxies that increases with increasing redshift; for H$_2$, this contribution is negligible 
at $z\lesssim 2$, but becomes systematically more important at higher redshifts, and for 
$\rho_{\rm SFR}$ they contribute less than $10$\% at $z\lesssim 3.5$, increasing slightly their contribution 
at higher redshifts; (ii) 
more massive galaxies {($M_{\rm stellar}>10^{10}\,M_{\odot}$)} are never important contributors to $\rho_{\rm HI}$, 
but host most of the H$_2$ 
in the universe at $z\lesssim 3$; in the case of $\rho_{\rm SFR}$, they make more than $30$\% of the 
total SFR at $z\lesssim 3.5$, percentage that slowly decreases at higher redshifts. The exact normalisations 
of $\rho_{\rm HI,m}$ and $\rho_{\rm H_2,m}$ slightly vary between the three different models. 
The largest differences 
obtained for the contribution to $\rho_{\rm HI}$, $\rho_{\rm H_2}$ and $\rho_{\rm SFR}$ from 
different selections in SFR are seen at $z\gtrsim 4$.

\section{Implications for the next generation of galaxy surveys}

In this section we discuss the implications of the results presented in $\S$~\ref{Contri} 
 for the current and next generation of galaxy surveys that aim to detect gas in galaxies. 
We first focus on the atomic content of galaxies in $\S$~\ref{HIgals} and then on the molecular gas in $\S$~\ref{H2gals}.

\subsection{Atomic hydrogen in high-redshift galaxies}\label{HIgals}
\begin{table*}
\begin{center}
\caption{The fraction of galaxies in the Lagos12 model that have { HI velocity-integrated fluxes 
larger than $3$ times the 1$\sigma$ sensitivity limits, $S_{\rm HI,lim}$ (column $1$), of 
 $4$ different surveys discussed in $\S$~$5.1$}, and that have stellar masses $<10^{8}\,M_{\odot}$, 
SFR$<0.1\,M_{\odot}\,\rm yr^{-1}$ or IR luminosities $<10^9\,L_{\odot}$, for different redshifts 
that cover the redshifts explored by the surveys (columns $2$ to $6$). 
The numbers in each column 
 are the fraction of galaxies fulfilling the { condition of a $>3\sigma$ detection at the given redshift, given the flux limit}, 
[$M_{\rm stellar}<M_{\star,\rm lim}$;~SFR$<$SFR$_{\rm lim}$;~$L_{\rm IR}<L_{\rm IR,lim}$], 
where $M_{\star,\rm lim}=10^{8}\,M_{\odot}$, SFR$_{\rm lim}=0.1\,M_{\odot}\,\rm yr^{-1}$ and 
$L_{\rm IR,lim}=10^9\,L_{\odot}$. {The sensitivity limits of the ASKAP surveys given in  (1) were calculated by 
\citet{Duffy12} for a bandwith of 
$3.86\,\rm km\,s^{-1}$, and for LADUMA by Sarah Blyth (priv. communication) for a bandwidth of $3.2\,\rm km\,s^{-1}$}.}\label{HIsurfs}
\begin{tabular}{ l c c c c c c }
\\[3pt]
\hline
(1) & (2) & (3) & (4) & (5) & (6) \\
\hline
1$\sigma$ sensitivity in ${\rm mJy\,km\,s^{-1}}$ & $z=0.08$ & $z=0.17$ & $z=0.5$ & $z=0.8$ & $z=1.1$  \\
\hline
 0.09 (DINGO DEEP) & [0.86;~0.96;~0.98] & [0.83;~0.94;~0.97]  & [0.34;~0.58;~0.88] & - & - \\
 0.2 (DINGO) & [0.85;~0.95;~0.98] & [0.75;~0.9;~0.97]    & [0.1;~0.27;~0.78] & - & - \\
1.592 (WALLABY) & [0.64;~0.85;~0.96] & [0.18;~0.54;~0.88]    & - &  - & - \\
0.106 (LADUMA BAND-L) & [0.86;~0.96;~0.98] & [0.83;~0.94;~0.97]  & [0.28;~0.53;~0.93] & [0.12;~0.25;~0.77]   & [0.02;~0.04;~0.6] \\ 
\hline
\end{tabular}
\end{center}
\end{table*}

There are two ways of studying HI, (i) through its emission at 
$21$~cm, either by individually detecting galaxies or stacking them to reach a high signal-to-noise ratio and 
(ii) through the absorption of the light of a background object by the presence of neutral hydrogen in the line-of-sight. 
We discuss future surveys in the context of HI in emission and absorption. 

{\it HI in emission.} Large blind surveys of the $21$~cm emission will be carried out by the 
next generation of radio telescopes, such as 
 the APERITIF project at the Westerbork Synthesis Radio Telescope,  
 the Australian SKA Pathfinder\footnote{\tt http://www.atnf.csiro.au/projects/mira/} (ASKAP),  
 the South-African SKA pathfinder MeerKAT\footnote{{\tt http://www.ska.ac.za/meerkat/}} (MeerKAT), 
 the Jansky Very Large Array\footnote{{\tt https://science.nrao.edu/facilities/vla}} (JVLA)
and in the 
longer term future, the Square
Kilometer Array\footnote{{\tt http://www.skatelescope.org/}} (SKA). 
One of the key surveys planned with ASKAP is DINGO, in which 
 galaxies up to $z\approx 0.43$, with the { 
%and spanning $7$ orders of magnitude in 
%stellar mass, $M_{\rm stellar}\sim 10^5-10^{12}\,M_{\odot}$, are expected to be detected in HI emission 
%\citep{Duffy12}. One of the 
key science goal of %DINGO is to 
understanding how the HI content of galaxies
co-evolves with other major galactic components, such as stars, dust and halo mass.} 
%(see also the CHILES survey at JVLA; \citealt{Fernandez13}). 
WALLABY will cover much larger areas in the sky, allowing the understanding of 
the effect of environment on the HI content of galaxies. 
 At higher redshifts, the Looking At the Distant Universe with the MeerKAT 
Array survey\footnote{\tt http://www.ast.uct.ac.za/laduma/Home.html} (LADUMA) plans to detect 
HI emission in individual galaxies up to $z\approx 1$. % extending the constraints on $\rho_{\rm HI,m}$ and $\rho_{\rm HI,SFR}$ to the more distant Universe.

Although these future surveys will provide unprecedented knowledge of the neutral gas in galaxies and its cycle, 
{ an important fraction of the HI in the universe is expected to be hosted 
in galaxies with small HI contents ($M_{\rm HI}<10^6\,M_{\odot}$) 
due to the steep low-mass end of the HI mass 
function.}
%we predict that most of the HI in the Universe that is locked up in galaxies
% with {low stellar masses ($M_{\rm stellar}<10^{9}\,M_{\odot}$) and 
% SFRs (SFR$<1\,M_{\odot}\,\rm yr^{-1}$;} see Fig.~\ref{denstotsgas} and Table~\ref{HIsurfs}).
These masses may even be too small to be detected 
 by ASKAP and MeerKAT. 
To overcome this problem and estimate the HI mass abundances in galaxies that are not individually
detected, it is possible to use  
intensity mapping or stacking of the $21$~cm emission line
 (see \citealt{Pritchard11} for a review). 
%Intensity mapping gives the two-dimensional spatial information at a given redshift; i.e. the auto-correlation 
%function of the $21$~cm emission line or cross-correlation with galaxy catalogues. Here, imaging 
% is obtained by aggregating the emission of thousands of galaxies on very large scales to get the summed signal 
%of galaxies that are not individually detected. 
The main difference in the estimates that can be made with intensity mapping is that what it is inferred is 
$\rho_{\rm HI}\cdot b_{\rm HI}$ (e.g. \citealt{Chang10}; \citealt{Masui13}), 
where $b_{\rm HI}$ is the bias in the clustering of HI selected sources 
 with respect to the expected underlying dark matter distribution, while 
stacking can infer a value for $\rho_{\rm HI}$ from a parent galaxy sample (e.g. \citealt{Delhaize13}; \citealt{Rhee13}).  
%Stacking has recently been applied 
%to estimate global HI densities at redshifts $z\sim 0.1-0.2$, which has shown to give estimates  
% in agreement with those inferred 
%from absorption-line measurements (e.g. \citealt{Delhaize13}; \citealt{Rhee13}). 
%On the other hand, $21$~cm intensity mapping from cross-correlations with 
%optical galaxy redshifts has been used to infer 
%the HI density at higher redshifts, $z\sim 0.8-1$ 
%and has also given measurements in agreement with those 
%inferred by absorption-line measurements (e.g. \citealt{Chang10}; \citealt{Masui13}).  
% The $21$~cm intensity mapping can even be used to measure the power spectrum directly without 
%the need of detecting individual galaxies. The latter can be done by using 
%the full information of the temperature, i.e. the $2$ dimensional projection on the sky 
%and the frequency information. If the power spectrum derived from 
%the intensity map is cross-correlated to galaxy catalogues, 
%it is possible to isolate the emission from a narrow redshift range and infer 
%information such as the HI content of galaxies below the detection limits of HI. This can place additional constraints 
%on our predictions of $\rho_{\rm HI,m}$ and $\rho_{\rm HI,SFR}$.

Another additional problem that these future surveys may encounter is the identification of 
UV, optical and/or IR counterparts. We argue that although to measure the total HI density at a 
given redshift is an important quantity for galaxy formation theory and cosmology, 
to be able to measure HI in galaxy populations which are selected by other means, either optical emission lines, 
broadband photometry, etc., is also valuable as additional constraints that can be put on galaxy formation 
models. However, even if a galaxy is detected in HI emission, its stellar mass 
and/or SFR may be too small to be measured, making the task of learning about galaxy formation from these 
HI surveys more challenging. 

{\it HI in absorption.} Another solution to the problem of faint HI emission is to observe neutral hydrogen in absorption. 
This has been extensively used to measure the number density of absorbers of different classes, such as 
DLAs, Lyman limit systems, etc. (see for example \citealt{Peroux03}). 
The first blind absorption survey in HI is going to be carried out by 
ASKAP\footnote{\tt http://www.physics.usyd.edu.au/sifa/Main/FLASH} (FLASH). 
This survey will allow measurements 
of $\rho_{\rm HI}$ as well as constraints on the evolution of faint HI galaxies by combining the FLASH survey with 
optical galaxy surveys as discussed above, in the redshift range $0.5<z<1$. 
%However, an important unknown in the conversion from the observed absorption strength to the HI column density, needed to 
%infer $\rho_{\rm HI}$, is 
%the spin temperatures of the absorbing gas. Assuming one spin temperature can 
%introduce a large uncertainty into the inferred $\rho_{\rm HI}$.
Recent HI absorption surveys have started to 
sample statistically the difference between absorbers around late- and early-type galaxies, as well as 
galaxies in different environments (\citealt{Tejos13}; \citealt{Tumlinson13}).  

We expect that only a combination of the surveys above (direct detection, absorption studies and stacking) will be able to 
 account for most of the HI in galaxies and determine the exact contribution to $\rho_{\rm HI}$ from galaxies with different 
properties. We summarise in Table~\ref{HIsurfs} the fraction of galaxies that have low stellar masses, SFRs or 
IR luminosities but that have {HI velocity-integrated fluxes that would be detected at a $>3\sigma$ level} 
by the deep HI surveys we described above. 
These numbers are intended to show the predicted 
frequency of HI detections that are likely to have no optical counterpart in large optical 
and near-IR galaxy surveys. A possible solution to this problem is to follow up these HI sources 
with deep optical and near-IR observations.
%, for instance with the Hubble Space Telescope, and in the future with 
%the James Webb Space Telescope\footnote{\tt http://www.jwst.nasa.gov/}. 

\subsection{Molecular hydrogen in high-redshift galaxies}\label{H2gals}

\begin{table*}
\begin{center}
\caption{The fraction of galaxies in the Lagos12 model that have {CO emission line velocity-integrated fluxes in 
the rotational transitions from 1-0 to 6-5 (column $1$) above a $3\sigma$ flux, with $1\sigma=1\,\rm mJy\, km\,s^{-1}$}, and that 
have stellar masses $<10^{9}\,M_{\odot}$, 
SFR$<10\,M_{\odot}\,\rm yr^{-1}$ or IR luminosities $<10^{10}\,L_{\odot}$, for different redshifts 
up to $z=2.8$ (columns $2$ to $7$). The numbers in each column are the fraction of galaxies at the given redshift fulfilling the conditions 
  [$M_{\rm stellar}<M_{\star,\rm lim}$;~SFR$<$SFR$_{\rm lim}$;~$L_{\rm IR}<L_{\rm IR,lim}$], 
where $M_{\star,\rm lim}=10^{9}\,M_{\odot}$, SFR$_{\rm lim}=10\,M_{\odot}\,\rm yr^{-1}$ and 
$L_{\rm IR,lim}=10^{10}\,L_{\odot}$.}
\label{H2surfs}
\begin{tabular}{ c c c c c c c }
\\[3pt]
\hline
(1) & (2) & (3) & (4) & (5) & (6) & (7) \\
\hline
CO transition & $z=0.1$ & $z=0.5$ & $z=1$ & $z=1.5$ & $z=2$ & $z=2.8$\\
\hline
CO(1-0) & [0.76;~0.99;~0.97]  &[0.12;~0.95;~0.81] & [0.0003;~0.71;~0.23]& [$10^{-5}$;~0.33;~0.004] & [$10^{-5}$;~0.06;~0]  & [0;~0;~0]    \\
CO(2-1) & [0.81;~0.99;~0.98]  &[0.5;~0.98;~0.9]   & [0.06;~0.86;~0.6]   & [0.01;~0.7;~0.29]  & [0.003;~0.4;~0.03]  & [0.0002;~0.03;~$10^{-5}$]    \\
CO(3-2) & [0.83;~0.99;~0.98]  &[0.68;~0.98;~0.93] & [0.22;~0.9;~0.71]   & [0.06;~0.78;~0.47]  & [0.06;~0.58;~0.14]  &  [0.02;~0.2;~0.0013]     \\
CO(4-3) & [0.82;~0.99;~0.98]  &[0.63;~0.98;~0.93]   & [0.27;~0.91;~0.73]  & [0.09;~0.8;~0.51]  & [0.04;~0.62;~0.2]  &  [0.008;~0.29;~0.004]    \\
CO(5-4) & [0.82;~0.99;~0.98]   &[0.57;~0.98;~0.91]    & [0.2;~0.9;~0.7]     & [0.06;~0.78;0.46] & [0.02;~0.58;~0.15] &  [0.006;~0.22;~0.002]   \\
CO(6-5) & [0.8;~0.99;~0.97]   &[0.41;~0.97;~0.88]    & [0.04;~0.85;~0.58] & [0.009;~0.68;0.26] & [0.003;~0.42;~0.03] &  [0.001;~0.04;~0.0006]   \\
CO(7-6) & [0.82;~0.99;~0.97]   &[0.3;~0.96;~0.86]    & [0.0006;~0.68;~0.22] & [0.0008;~0.33;0.005] & [0.002;~0.07;~0.0003] &  [0.006;~0;~0.0003]   \\
\hline
\end{tabular}
\end{center}
\end{table*}

To trace molecular gas, lower order molecule transitions are 
better, as they are most commonly excited predominantly through collisions (in optically thick media), and they are more easily 
excited than the higher order molecule transitions. The most used molecule to trace H$_2$ is CO, which is the most 
abundant molecule after H$_2$. However, the condition of optically thick media, that defines the excitation means, 
may break-down in very low metallicity gas, where the CO lines 
 become optically thin (see \citealt{Carilli13} for a recent review). 
Lagos12 find that the number of galaxies with gas metallicities $Z_{\rm gas}<0.01\,Z_{\odot}$ 
is $<0.05$\% for galaxies with IR luminosities $>10^9\,L_{\odot}$ in the redshift range $0\le z\le 6$.
This IR luminosity limit corresponds roughly  
to a SFR$\sim 0.1\,M_{\odot}\,\rm yr^{-1}$. Thus, our model indicates that CO can be used as a good tracer 
of H$_2$ for the galaxies that dominate the H$_2$ content of the universe, which are those with 
SFR$>1\,M_{\odot}\,\rm yr^{-1}$ or with $M_{\rm stellar}>10^{10}\,M_{\odot}$. 
Galaxies in which CO is not a good tracer of H$_2$ (i.e. those with low gas metallicities) represent only a small
 contribution to the H$_2$ content of the universe. The downside of CO is the CO-H$_2$ conversion factor, which is a strong 
function of metallicity and other physical conditions in the ISM of galaxies \citep{Boselli02}. The positive 
side is that the CO-H$_2$ conversion factor has started to be systematically explored 
in cosmological simulations of galaxy formation with the aim of informing how much variation with 
redshift is expected (\citealt{Narayanan12}; \citealt{Lagos12}; \citealt{Popping13b}).
%There are other tracers of 
%H$_2$ that are free of strongly varying conversion factors, and that can cleanly trace H$_2$ down to small 
%optical depths. An example of this are the fluorine-bearing molecules \citep{Neufeld05}. These have the downside 
% that their abundance is four orders of magnitude smaller than that of CO. In this section we therefore 
%focus on CO as the primary tracer of H$_2$. 

Blind molecular emission surveys are the best way of studying the cold gas content 
of galaxies in an unbiased way. 
%An advantage of using molecule emission is that the 
%centrifugal force stretches the bond as the $J$ transition increases, allowing 
% the redshift to be unequivocally measured from just two transitions (e.g. \citealt{Weiss09}). 
However, they require significant time integrations to ensure 
good signal-to-noise. Alternatively, molecular emission line surveys can be guided by 
well-known UV, optical, IR and/or radio surveys, with well identified positions and spectroscopic 
or photometric redshifts. Even if only highly star-forming galaxies are used as guide for follow-up 
molecular emission line surveys, it is possible to place strong constraints on models of galaxy formation, as we 
show in $\S$~\ref{Contri}, and to provide new insight into the co-evolution of the different 
baryonic components of galaxies.  
 
The 
%currently available band~3 ($84-116$~GHz) in the 
Atacama Large Millimeter Observatory (ALMA) 
can currently detect CO(1-0) and CO(2-1) up to $z\approx 0.84$.
%covers the lowest CO transition, 1-0, at $z\lesssim 0.37$.
%, while 
%higher CO transitions are detectable in this band only at $z\gtrsim 0.98$. 
%ALMA band~4, which will be available at 
%cycle~2, will cover the next CO transition, the 2-1, in the redshift range $0.41-0.84$. 
%The other available 
%bands (6, 8 and 9) cover higher CO transitions. 
%In the future, the availability of band~1 and band~2
%in ALMA will allow to observe CO$(1-0)$ in the redshift ranges $1.5\lesssim z\lesssim 2.7$ and 
%$0.3\lesssim z\lesssim 0.8$, respectively.
At $z\gtrsim 1.3$, instruments that are primarily designed to detect atomic hydrogen, such as  
 the JVLA are able to detect  
the emission from the redshifted frequencies of low-J
CO transitions.%, while in the future MeerKAT and SKA will be able to do the same 
%for fainter galaxies.
 A good example of this is presented in \citet{Aravena12}, who performed a JVLA blind survey 
in a candidate cluster at $z\sim 1.5$. Aravena et al. obtained two detections and were 
able to place constraints on the number density of bright CO($1-0$) galaxies. Similarly, blind CO surveys in 
small areas of the sky are ongoing using the JVLA 
and the Plateau de Bureau Interferometer \citep{Walter14}. 
%These surveys 
%aim to place constraints on $\rho_{\rm H_2}$ for galaxies with molecular emission lines bright enough to be detected 
%in these surveys. 

For faint sources, the intensity mapping technique described in $\S$~\ref{HIgals} 
can also be applied to molecular emission. 
%In this case the main challenges come from 
%subtracting the foreground signal which is typically more than an order of magnitude brighter than 
%the high redshift signal, and from line confusion from galaxies at different redshifts. 
The line confusion from galaxies at different redshifts 
%The former 
%can be done cleanly if the foreground continuum is smooth, while the latter 
can be overcome by 
cross-correlating different molecular and/or atomic emission lines. Different emission lines coming from the same galaxies 
would show a strong correlation, while emission lines coming from galaxies at different redshift would not be correlated. 
% Thus, cross-correlating can isolate a narrow redshift range. 
From this, the quantity that is derived 
is the two-point correlation function or power spectrum weighted by the total emission in the spectral lines being correlated 
(see \citealt{Pritchard11} for a review).
% From this, it is possible to infer 
%the time evolution of the total emission of a molecular or atomic emission line. 
The power of using intensity mapping for CO lines have been shown and discussed for instance by \citet{Carilli11} and 
\citet{Gong11}. Carilli argues that intensity mapping of CO lines 
can be performed with telescopes of small field-of-view at redshifts close or at 
the epoch of reionisation ($z\sim 3-8$). 

Unlike in HI, optical counterparts could be easily identified in the case of individual detection of molecular emission, 
 since they are expected to be bright and have large stellar masses.
  Table~\ref{H2surfs} shows the fraction of galaxies in model samples selected by their {CO velocity-integrated 
flux and} that have stellar masses $<10^9\,M_{\odot}$, SFR$<10\,M_{\odot}\,\rm yr^{-1}$ or IR luminosities 
$<10^{10}\,L_{\odot}$. These fractions are encouraging as they are generally low compared to HI mass selected samples, 
particularly at high redshifts; this means that we expect effectively all of the molecular emission lines detected to have 
detectable optical/IR counterparts. From this, one can conclude that the most efficient strategy to uncover the 
H$_2$ in the universe is the follow-up of already existing optical/IR/radio catalogues with submillimeter and/or radio telescopes 
with the aim of detecting molecular emission lines.  

\nocite{Combes11}
\nocite{Combes13}

\section{Conclusions}\label{conclusion} 

We have presented predictions for the contribution to the densities of atomic and molecular hydrogen 
from galaxies with different stellar masses, SFRs and IR luminosities in the context of 
galaxy formation in a $\Lambda$CDM framework. 
We use three flavours of the {\tt GALFORM} semi-analytic model of galaxy formation, the Lagos12, Gonzalez-Perez14 and Lacey14 models. 
For quiescent star formation, 
the three models use the pressure-based SF law of \citet{Blitz06}, in which the ratio between the 
gas surface density of H$_2$ and HI is derived from the radial profile of the hydrostatic pressure of the disk, and calculates 
the SFR from the surface density of H$_2$. The advantage of this SF law is that the atomic and molecular gas phases 
of the ISM of galaxies are explicitly distinguished. Other physical processes in the three models are different, such 
as the adopted IMF and the strength of both the SNe and the AGN feedback, as well as the cosmological parameters adopted.

We identify the following trends in the three models and regard them as robust:  

(i) The HI density shows almost no evolution at $z\lesssim 2$, and slowly decreases with increasing 
redshift for $z\gtrsim 2$. 

(ii) The H$_2$ and the SFR densities increase strongly from $z=0$ up to $z\approx 2-3$, followed by a slow 
decline at higher redshifts. The models predict an increase in SFR larger than for H$_2$ due to the contribution from starbursts, 
which is larger for the SFR (close to $50$\% in the three models at $z\approx 3$) than for  
 H$_2$ (less than $10$\% at $z\approx 3$). The latter is due to starbursts being on average more efficient at converting 
gas into stars, and therefore contributing more to the SFR density than the gas density of the universe. 

(iii) The three models predict densities of SFR, HI and H$_2$ in broad agreement with the observations 
at $z\lesssim 3$. A higher redshifts important deviations are observed between the predicted $\rho_{\rm HI}$ and 
the observationally inferred values in the three models. We argue that this is not unexpected as the HI 
in our simulations correspond exclusively to HI in galaxies and no contribution from the HI in the IGM 
is included. The latter is expected to become important at $z\gtrsim 3$ \citep{VanDeVoort12}. 

(iv) The density of HI is always dominated by galaxies with low SFR (SFR$<1\,M_{\odot}\,\rm yr^{-1}$), and low stellar masses   
($M_{\rm stellar}<10^9\,M_{\odot}$), while the H$_2$ density is dominated by galaxies with relatively large SFRs  
 (SFR$>1\,M_{\odot}\,\rm yr^{-1}$) and large stellar masses ($M_{\rm stellar}>10^{10}\,M_{\odot}$). The latter is also 
true for the global SFR density. 

(v) The predicted evolution of the global SFR density observed in the universe can be largely explained as driven by the  
steep decline of the molecular mass towards $z=0$. The combination of the evolution of the 
total neutral gas, atomic and molecular, and the increasing galaxy sizes with decreasing redshift 
explain the evolution of the H$_2$ density.
 The global SFR density evolution can therefore be linked to the evolution of the neutral gas 
surface density of the galaxies dominating the SFR in the Universe at a given time. 
The evolution of the neutral gas content of galaxies is set by the balance between inflows and outflows 
of gas in galaxies, and therefore also plays a key role in the evolution of the molecular gas, that ultimately set 
the SFR. Thus, one must understand that the H$_2$ content and SFR of galaxies in the star-forming sequence 
  of the SFR-stellar mass plane are set by the self-regulation of inflow and outflows and that small { deviations}
to this self-regulation produce the time evolution of this sequence. 

(vi) We find that a group of observations can be connected and understood in the
 models, including the H$\alpha$ and UV luminosity functions, 
 distribution function of the global SFR density on stellar mass, the global SFR density from highly star-forming galaxies, the 
inferred HI density from massive galaxies in the local Universe, and the inferred H$_2$ density of highly star-forming galaxies 
at high-redshifts. This broad set of comparisons between observations and model predictions that 
test at the same time the stellar mass and SFR and the  
neutral gas content of galaxies is unprecedented in semi-analytic models of galaxy formation.

We discuss our findings in the context of future surveys and suggest that optical identification of 
faint HI counterparts is going to be difficult due to the low stellar masses and SFRs we expect for them. 
These faint HI galaxies are, however, key to our understanding of galaxy formation as we expect them to 
dominate the HI content of the universe. 
 We suggest that stacking techniques are promising to measure the contribution to HI 
from galaxies selected by their stellar mass or SF activity. Such measurements have the potential to place strong constraints on 
galaxy formation models. Regarding H$_2$, the easiest way to make progress is to use current surveys with well identified   
optical, IR and/or radio positions and well known spectroscopic or photometric redshifts and follow them up 
with current facilities, with the downside that the selection function becomes more complex to model.  
The fact that most of the H$_2$ is predicted by our model to be hosted by relatively massive, star-forming galaxies, which are 
already detectable by optical and IR surveys, 
it is implied that follow-up surveys will be able to uncover the H$_2$ density of the universe without the need   
of blind surveys. From such observations, it is implied that both direct detection and stacking would provide valuable information 
on the cold gas content of galaxies and place strong constraints on galaxy formation models. 

\section*{Acknowledgements}

We thank Michelle Furlong, Alvaro Orsi, Madusha Gunawardhana, Benjamin Magnelli and David Sobral, Andrew Baker and Sarah 
Blyth for providing data compilations for this work. We also thank David Sobral, Gergo Popping, Danail Obreschkow, Andrew Hopkins, Peder 
Norberg, Tim Davis and Izaskun Jimenez-Serra 
for the discussions that helped to improve this work. We thank the anonymous referee for helpful suggestions that improved this work. 
 We thank Peder Norberg, John Helly and Alex Merson  
 for running the three models used in this paper for the Millennium database and for creating lightcones with them.
The research leading to these results has received funding from 
the European Community's Seventh
Framework Programme ($/$FP7$/$2007-2013$/$) under grant agreement no 229517.
This work used the DiRAC Data Centric system at Durham University, operated by the Institute for Computational Cosmology on behalf of the STFC DiRAC HPC Facility ({\tt www.dirac.ac.uk}). 
This equipment was funded by BIS National E-infrastructure capital grant ST/K00042X/1, STFC capital grant ST/H008519/1, and STFC DiRAC Operations grant ST/K003267/1 and Durham University. DiRAC is part of the National E-Infrastructure.

%----------------------------------------------
\bibliographystyle{mn2e_trunc8}
\bibliography{GlobalHIH2}
%---------------------------------------------------------------------

\label{lastpage}
\appendix
\section[]{Scaling stellar masses and star formation rates to our adopted IMF}\label{IMFs}

Due to our chose of the \citet{Kennicutt83} IMF, we have to scale
the observational inferences of stellar masses and SFRs made with other choices of 
IMF. The other choices of IMFs in the compilation of data presented in Fig.~\ref{denstots} and Fig.~\ref{denstots2} 
 are the \citet{Chabrier03} IMF, the \citet{Salpeter55} IMF and the 
\citet{Baldry03} IMF. 

For stellar masses, we follow the conversions in Table~\ref{SMsIMF} that have been taken from 
\citet{Bell03} and \citet{Gilbank10}.

%First, we define the four IMFs involved in these conversions. The distribution of the masses of stars 
%formed in one episode follow ${\rm d}N(m)/{\rm d\, ln}\,m\equiv m^{-x}$, where N is the number of stars of mass $m$ formed, 
% and $x$ is the IMF slope. For a \citet{Kennicutt83} IMF, $x=1.5$ for masses in the range $1\,M_{\odot}\le m\le 100\,M_{\odot}$ and 
%$x=0.4$ for masses $m< 1\,M_{\odot}$. For a \citet{Salpeter55} IMF, $x=1.35$ for masses in the range 
%$0.1\,M_{\odot}\le m\le 100\,M_{\odot}$. For a \citet{Chabrier03}, $x=1.3$ for masses in the range $1\,M_{\odot}\le m\le 100\,M_{\odot}$
% and has a functional form 

\begin{table}
\begin{center}
\caption{Scaling of the stellar mass from different IMF choices to a 
\citet{Kennicutt83} IMF. The conversion given for each IMF follows 
$M^{\rm Kenn}_{\rm stellar}={\rm corr}\,\cdot M^{\rm IMF}_{\rm stellar}$, where 
$M^{\rm Kenn}_{\rm stellar}$ is the stellar mass inferred for a \citet{Kennicutt83} IMF and 
$M^{\rm IMF}_{\rm stellar}$ the stellar mass inferred in observations that adopted a different IMF. 
These are taken from the work of \citet{Bell03} and \citet{Gilbank10}.}\label{SMsIMF}
\begin{tabular}{l c }
\\[3pt]
\hline
IMF & corr \\
\hline
\citet{Salpeter55} & 0.5\\
\citet{Chabrier03} & 0.89\\
\citet{Baldry03} & 2.25\\
\hline
\end{tabular}
\end{center}
\end{table}

The case of SFRs is more complex than the stellar mass, since the conversion depends on the SFR tracer used. 
We here follow the scalings calculated by Gonzalez-Perez et al. (2013, in prep.). Gonzalez-Perez et al. use 
the {\tt PEGASE} \citep{Fioc99}, and assume a constant solar etallicity and SFR to calculate the H$\alpha$, H$\beta$, 
[OII], FUV and radio luminosities for each choice of IMF. Note, however, that the conversions are subject to 
systematics related to the assumptions of SF history and metallicity (e.g. \citealt{Wilkins12}).
 We summarise in Table~\ref{SFRIMF} the scalings applied to each observationally inferred 
SFR depending on the adopted IMF and used tracer to convert to a \citet{Kennicutt83} IMF. Here, FUV corresponds to a filter 
centered at $1500$\AA~and width $400$\AA~and NUV to a filter centered at $2000$\AA~and width $400$\AA.

\begin{table}
\begin{center}
\caption{Scaling to convert the observationally inferred SFR that adopted an IMF different than 
the \citet{Kennicutt83} IMF.
The conversion given for each IMF follows 
${\rm SFR}^{\rm Kenn}={\rm corr}\,\cdot {\rm SFR}^{\rm IMF}$, where 
${\rm SFR}^{\rm Kenn}$ would be the SFR inferred for a \citet{Kennicutt83} IMF and 
${\rm SFR}^{\rm IMF}$ the SFR inferred in observations that adopted a different IMF. Each column shows a different 
SFR tracer.}\label{SFRIMF}
\begin{tabular}{l c c c c}
\\[3pt]
\hline
IMF & optical E.L. &  1500\AA& 2000\AA \\
\hline
\citet{Salpeter55} & 0.93     &0.79 & 0.77    \\
\citet{Chabrier03} & 1.57     &1.26 & 1.22    \\
\citet{Baldry03} &    2.26       &1.56 & 1.63      \\
Top-heavy IMF ($x=1$) & 3.132 &1.892 & 1.772  \\
\hline
\hline
IMF & 2500\AA& 2800\AA & 3550\AA\\
\hline
\citet{Salpeter55}    &  0.76  & 0.747 & 0.717\\
\citet{Chabrier03}    &  1.203 & 1.172 & 1.11\\
\citet{Baldry03}      &  1.445 & 1.385 & 1.266\\
Top-heavy IMF ($x=1$) &  1.771 & 1.616 & 1.434\\
\hline
\hline
IMF & FIR & radio\\
\hline
\citet{Salpeter55}    &  0.807 & 0.77\\
\citet{Chabrier03}    &  1.296 & 1.22\\
\citet{Baldry03}      &  1.636 & 1.46\\
Top-heavy IMF ($x=1$) &  2.022 & 1.678\\
\hline
\end{tabular}
\end{center}
\end{table}

\section[]{The local universe HI and H$_2$ mass functions}\label{MFs}

In Fig.~\ref{HIcompall} we show the predicted HI and H$_2$ mass functions at $z=0$ for the Lagos12, Lacey14 and 
Gonzalez-Perez14 models. We compare with observations of the HI mass function from \citet{Zwaan05} and \citet{Martin10}, 
and of the inferred H$_2$ mass function from \citet{Keres03}. The latter observations correspond to CO($1-0$) molecular emission line 
in galaxies that were previously selected from their emission in the $B$-band or $60\mu$m. We then use a Milky-Way type 
conversion between the CO($1-0$) emission and the H$_2$ mass, and infer a H$_2$ mass function. We showed in \citet{Lagos12} 
that doing the opposite exercise of converting predicted H$_2$ masses to CO($1-0$) emission using 
radiative transfer, photon dominated region codes to calculate the CO-H$_2$ conversion factor, 
leads to similar CO($1-0$) luminosity function than the one obtained by assuming a constant conversion factor. This is due 
to the fact that most model galaxies at $z=0$, which have CO($1-0$) luminosities above the observational limit of
\citet{Keres03}, have gas metallicities and SFR densities that are not too different from normal spiral galaxies. 
The three models give predictions that are in good agreement with the observations. 

\begin{figure}
\begin{center}
\includegraphics[trim = 1.5mm 0.5mm 1mm 1mm,clip,width=0.48\textwidth]{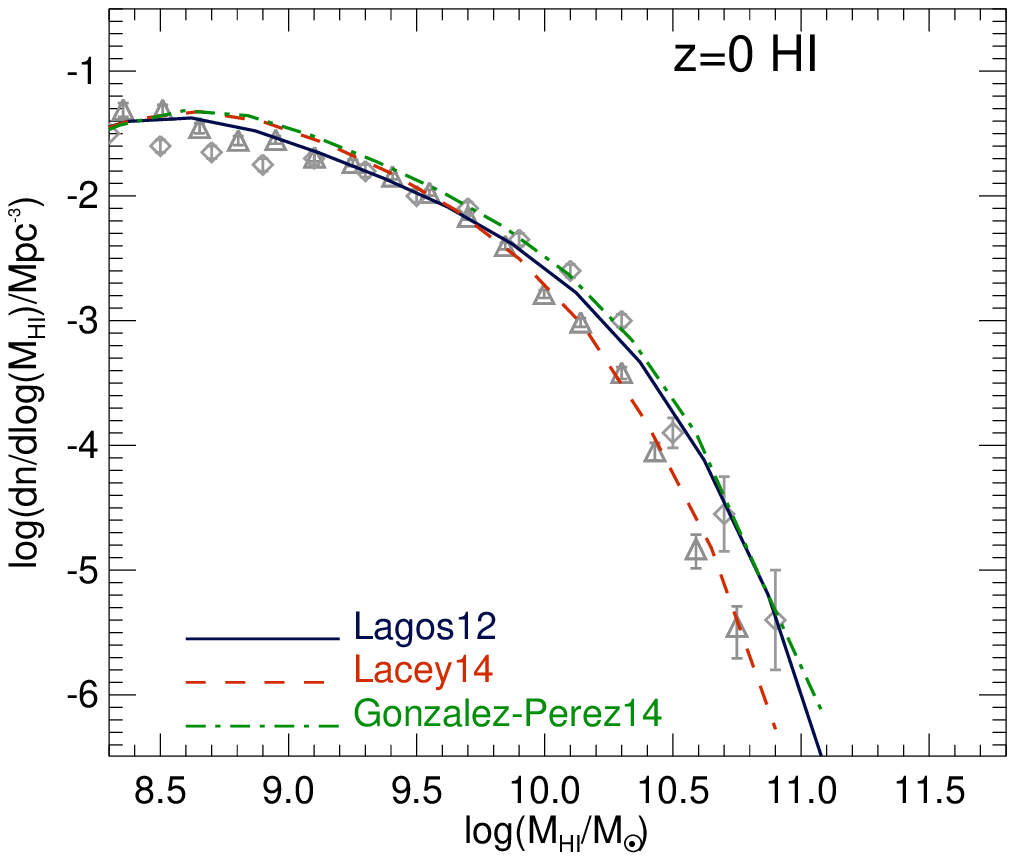}
\includegraphics[trim = 1.5mm 0.5mm 1mm 1mm,clip,width=0.48\textwidth]{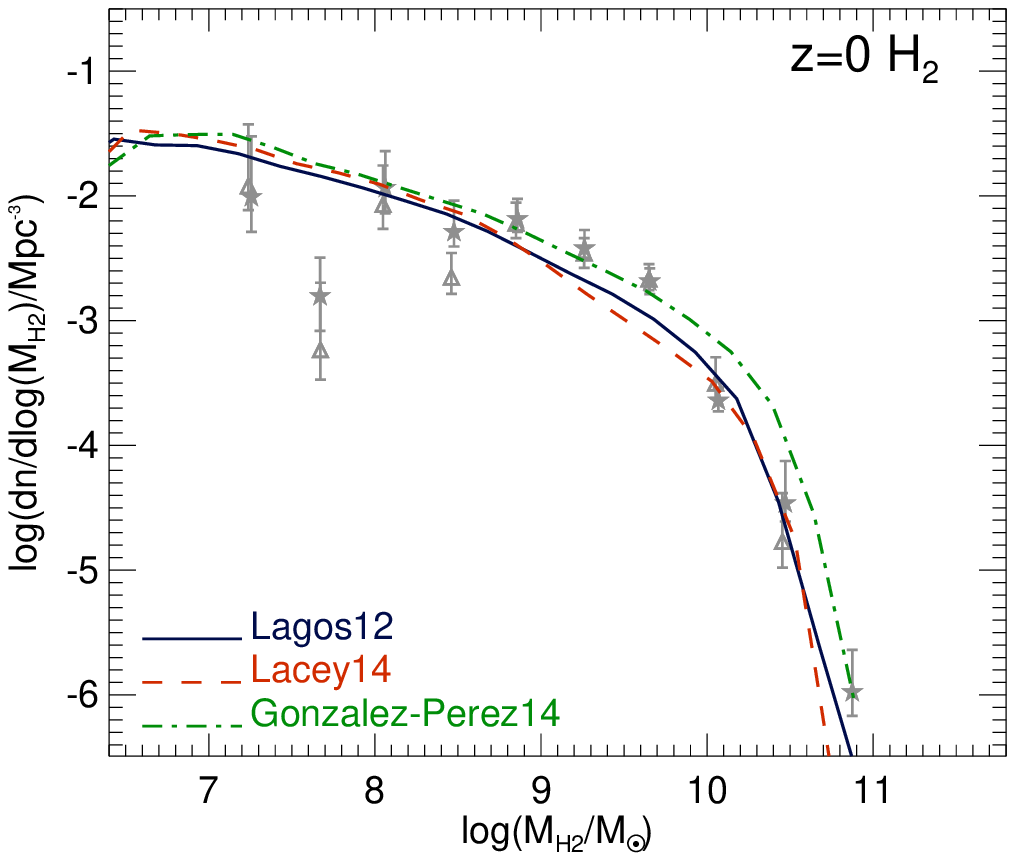}
\caption{{\it Top panel:} The HI mass function at $z=0$ for the Lagos12, Lacey14 and Gonzalez-Perez14 models, as labelled. 
Symbols show observational results at $z=0$ from \citet{Zwaan05}
using HIPASS (diamonds) and \citet{Martin10} using ALFALFA (squares). 
{\it Bottom panel:} Same as the top panel but for H$_2$. Here we show observational inferences from \citet{Keres03}
 using direct detection of the $\rm CO(1-0)$ emission line in 
$B$-band (triangles) and $60\,\mu$m (filled circles) selected samples of galaxies.
 The observational mass function is then calculated assuming the Milky Way $\rm H_2$-to-CO conversion factor, 
$N_{\rm H_2}/\rm cm^{-2}=2\times10^{-20}\, I_{\rm CO}/\rm K\,km\,s^{-1}$. 
 Here $N_{\rm H_2}$ is the column density of H$_2$ and
$I_{\rm CO}$ is the integrated CO$(1-0)$ line intensity per unit
surface area.}
\label{HIcompall}
\end{center}
\end{figure}

\section[]{The dark matter halos of the Lagos12, Lacey14 and Gonzalez-Perez14 models}\label{Resolution}

\subsection{The Lagos12, Lacey14 and Gonzalez-Perez14 models in the Millennium database}

The predictions of the three {\tt GALFORM} models used in this paper will be publicly available from 
the Millennium database. However, the reader should keep in mind that the runs performed in this paper are not `standard' 
in the sense that we use Monte-Carlo realisations of a wider dynamical range of halo masses compared to the $N$-body 
simulations Millennium \citep{Springel05}, used by the Lagos12 model, and 
 MS-W7 (same as the Millennium simulation but with WMAP7 cosmology), used 
by Gonzalez-Perez14 and Lacey14. This means that the runs shown in this paper follow the formation of galaxies 
in dark matter halos that extend to much lower masses than the two $N$-body simulations above. The minimum 
halo mass of the Millennium and MS-W7 simulations is fixed on $1.72\times 10^{10}\,h^{-1}\,M_{\odot}$. 
This has the limitation that the results presented here for the HI density cannot 
be fully reproduced with the galaxies in the Millennium database. This is due to the faint nature of the galaxies that dominate 
$\rho_{\rm HI}$, which are hosted by dark matter halos that are not resolved in the Millennium and 
MS-W7 $N$-body simulations. All the results related to H$_2$ and SFR density (including the H$\alpha$ and UV luminosity functions) 
can be reproduced using the galaxies in the Millennium database at $z<4$. This is because the galaxies that dominate 
these statistical measurements are relatively massive, and hosted by halos that are well resolved in the $N$-body simulations. 
At $z>4$, the SFR density is not fully resolved in the resolution of the Millennium and MS-W7 $N$-body simulations. 

We have also performed runs using the dark matter halo merger trees of the Millennium-II simulation, which 
has a minimum halo mass fixed at $1.38\times 10^8\,h^{-1}\,M_{\odot}$ \citep{Boylan-Kolchin09}. We find that 
the dynamical range of the Millennium-II is sufficient to resolve the HI in galaxies at $z\lesssim 2.5$. 
However, at higher redshifts, 
the Millennium-II misses the lowest halo masses that still contribute to the HI in the universe ($10^7-10^8\,h^{-1}\,M_{\odot}$), that 
are covered by our Monte-Carlo merger trees thanks to the redshift scaling applied to the minimum halo mass. 
  
\subsection{Resolution tests}

\begin{table}
\begin{center}
\caption{Simulations used to test the convergence of the predictions of the models. 
Column (1) gives the name of each simulation, (2) the minimum halo mass adopted, $M_{\rm halo,min}$, and 
(3) the scaling of the minimum halo mass with redshift. The standard simulation used throughout the manuscript is r3.}\label{ResMasses}
\begin{tabular}{c l l }
\\[3pt]
\hline
Simulation & $M_{\rm halo,min}/M_{\odot}\,h^{-1}$ & Scaling\\
\hline
r1 & $10^{10}$ & $(1+z)^{-3}$\\
r2 & $5\times10^{9}$ & $(1+z)^{-3}$\\
r3 & $5\times10^{8}$ & $(1+z)^{-3}$\\
r4 & $2.5\times10^{8}$& $(1+z)^{-3}$\\
r5 & $10^{10}$& no scaling\\
r6 & $5\times 10^{9}$ & no scaling\\
\hline
\end{tabular}
\end{center}
\end{table}

A questions that comes immediately from the lack of resolution in the Millennium and MS-W7 $N$-body simulations is whether the 
results presented here for the HI density are converged or not. To answer this question, 
we performed a simple exercise: we run the same model 
under $6$ different resolutions, with and without the scaling with redshift of the minimum halo mass. 
Table~\ref{ResMasses} summarises the different runs.
The scaling with redshift is also changed to study 
%It is interesting to show 
the effect of this scaling on the HI density to reproduce the conditions of 
the $N$-body runs (i.e. fixed minimum halo mass) and also to show the importance of adding the redshift scaling 
in the Monte-Carlo merger trees. 
%With in mind, we run the Lagos12 model with no minimum halo mass scaling 
%with redshift and using $M_{\rm  halo,min}=10^{10}\,h^{-1}\,M_{\odot}$ (refer to as `r5') and 
%$M_{\rm  halo,min}=5\times 10^{9}\,h^{-1}\,M_{\odot}$ (refer to as `r6'). 
Fig.~\ref{ResRuns} shows a visualization of the different resolutions of all the 
runs. We also show in Fig.~\ref{ResRuns} the resolution of the Millennium I and Millennium 
II $N$-body simulations.   

\begin{figure}
\begin{center}
\includegraphics[trim = 1.5mm 2.0mm 1mm 5mm,clip,width=0.47\textwidth]{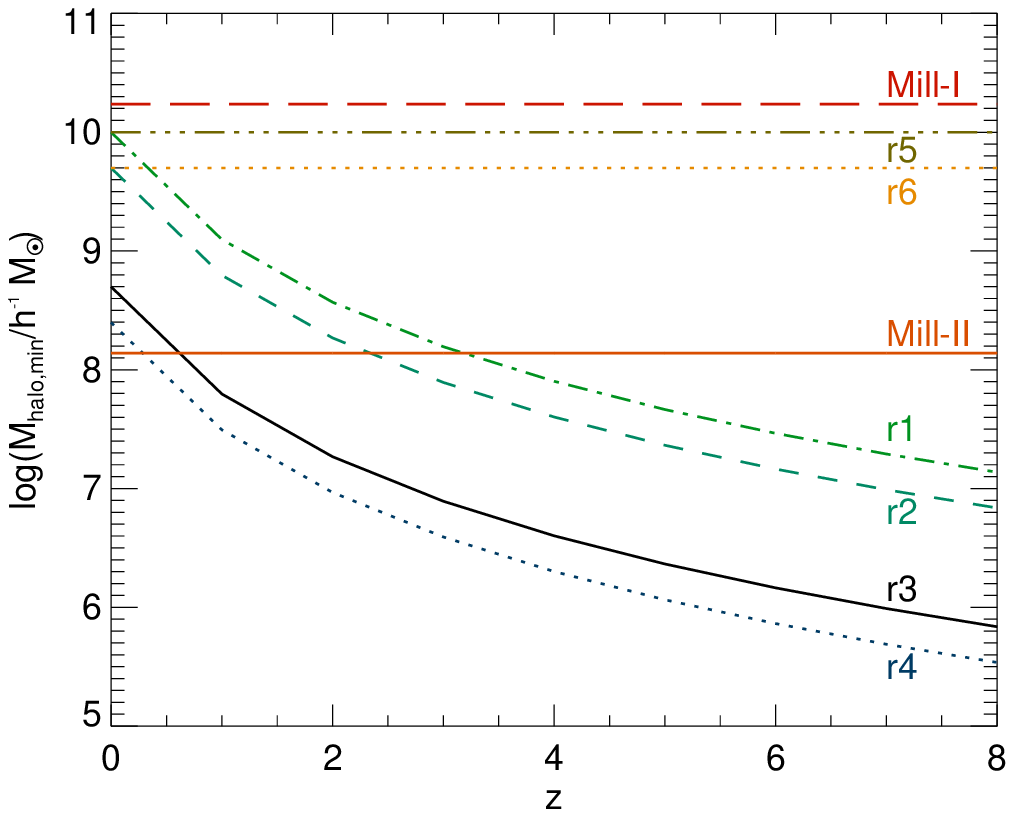}
\caption{The minimum halo mass in the different resolution runs as a function of redshift. 
The lines correspond to the r1 (dot-dashed line), r2 (dashed line), r3 (solid line), r4 (dotted line), r5 (horizontal triple dot-dashed line) 
and r6 (horizontal dotted line) runs. As a reference, we also show 
the resolution of the Millennium I (horizontal long dashed line) and Millennium II (horizontal solid line) simulations.}
\label{ResRuns}
\end{center}
\end{figure}

\begin{figure}
\begin{center}
\includegraphics[trim = 1.5mm 1.5mm 1mm 3mm,clip,width=0.45\textwidth]{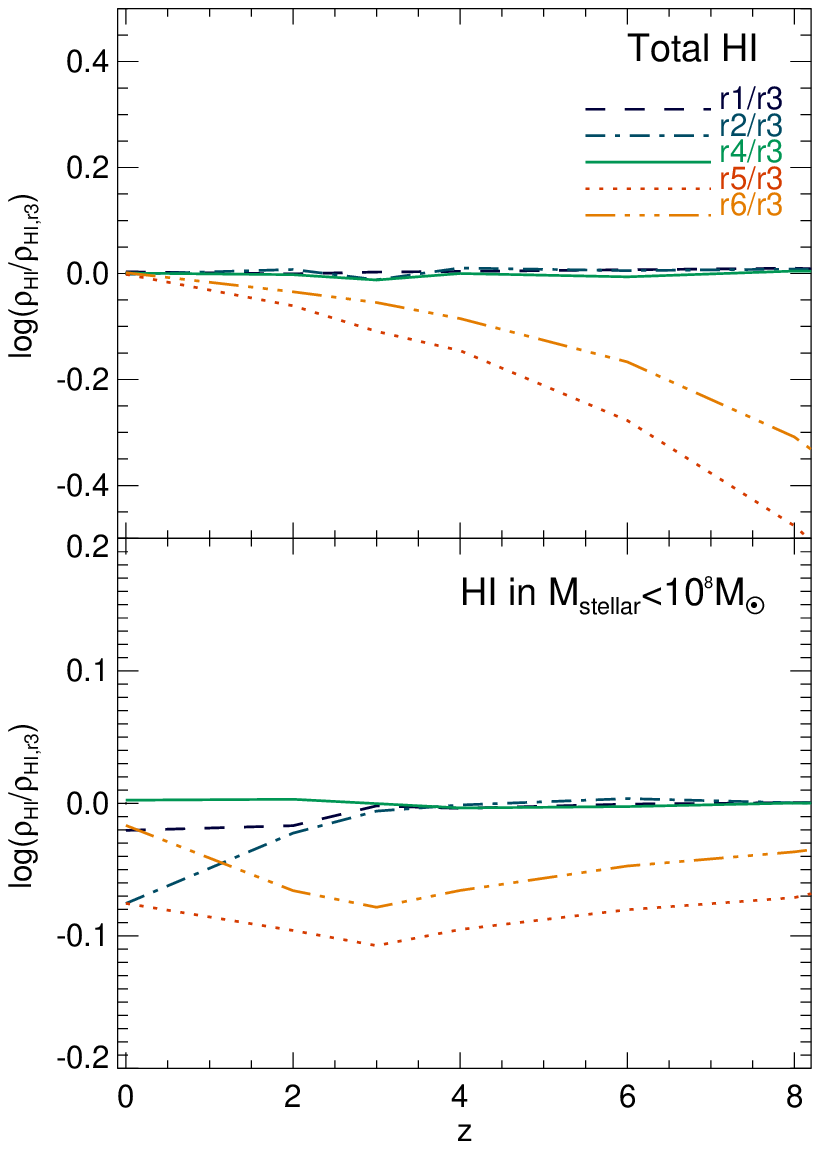}
\caption{The ratio between the HI density in a given resolution and the r3 run, as a function of redshift. The top panel shows 
the global HI density ratio and the bottom panel shows the HI density in galaxies with $M_{\rm stellar}<10^8\,M_{\odot}$ in the 
Lagos12 model. Each line shows a different resolution test as labelled with respect to our standard run, the r3 resolution.}
\label{ResTest}
\end{center}
\end{figure}

\begin{figure}
\begin{center}
\includegraphics[width=0.45\textwidth]{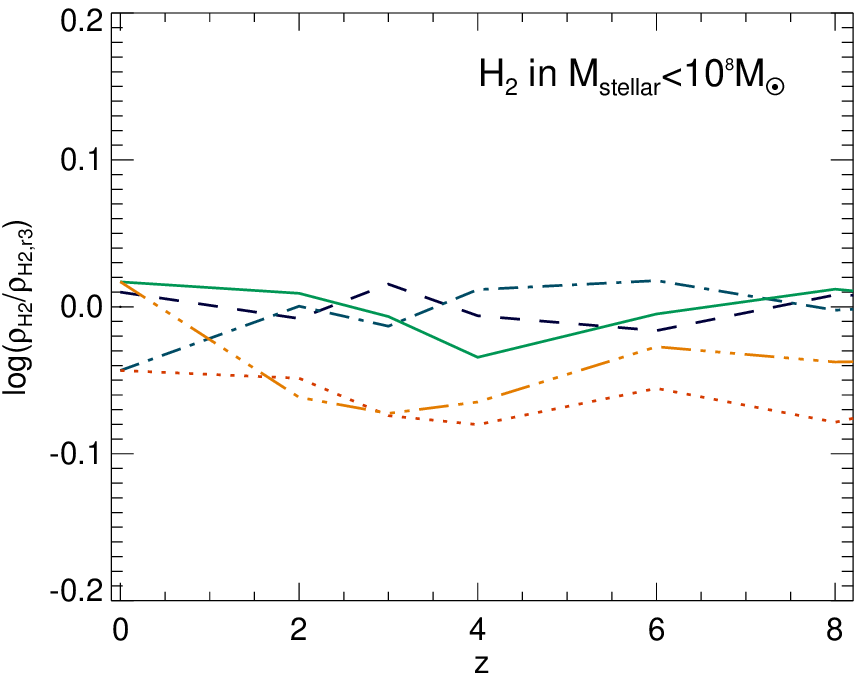}
\caption{Same as the bottom panel of Fig.~\ref{ResTest} but for H$_2$.}
\label{ResTest2}
\end{center}
\end{figure}

We show in Fig.~\ref{ResTest} the global HI density and the 
contribution to it from galaxies with $M_{\rm stellar}<10^8\,M_{\odot}$ for the Lagos12 model in the five Monte-Carlo tree 
resolutions, all normalised to our standard choice, the `r3' Monte-Carlo merger trees. By comparing the 
r3 with the r4 resolution, we can conclude that our 
standard resolution (r3) is converged to a factor better than $2$\%. When comparing the resolutions r1 and r2 with our standard r3 
resolution, we conclude that although the global HI density is converged to a factor better than $10$\%, the contribution from 
low mass galaxies is not. This is shown in the bottom panel of Fig.~\ref{ResTest}, where r1 and r2 predict lower 
contributions from the low mass galaxies to $\rho_{\rm HI}$ by $\approx 60$\% and $\approx 10$\%, respectively. The importance 
of the redshift scaling introduced in the minimum halo mass in our standard run is crucial to make our predictions converge. 
This can be concluded from the large offset between the r3 and the runs r5 and r6, which increases with increasing redshift. 
\citet{Lagos11} argue that this is the reason why previous models predict a monotonically increasing 
H$_2/$HI global density ratio with increasing redshift (e.g. \citealt{Obreschkow09}; \citealt{Popping13}), and 
miss the turnover in this ratio, which takes place at $z\approx 5$ (i.e. in the regime where the fixed resolution run 
largely underestimates the global HI density). Also note that the runs r5 ad r6 predict a $\rho_{\rm HI}$ that 
converges to the high-resolution $\rho_{\rm HI}$ at $z\lesssim 1$. {The contribution from low mass galaxies 
($M_{\rm stellar}<10^8\,M_{\odot}$) to the density of H$_2$, shown in Fig.~\ref{ResTest2} is converged to better than $10$\% 
at the resolution of r2, meaning that the contribution from small galaxies is resolved in or standard simulation. 
The fact that the convergence is achieved 
at lower resolutions compared to the case of HI is due to the flatter low-mass end of the H$_2$ mass function compared to the steepness 
of the low-mass end of the HI mass function.}

\end{document}